\documentclass[twocolumn,aps, prl, showpacs, preprintnumbers, amsmath, amssymb, a4paper, superscriptaddress]{revtex4-1}

\usepackage{graphicx}
\usepackage{dcolumn}
\usepackage{bm}
\usepackage{epstopdf}
\usepackage[colorlinks=true,citecolor=black,linkcolor=black,urlcolor=blue,anchorcolor=black]{hyperref}
\usepackage{color}

\begin{document}
\newcommand{\Arg}[1]{\mbox{Arg}\left[#1\right]}
\newcommand{\bb}{\mathbf}
\newcommand{\braopket}[3]{\left \langle #1\right| \hat #2 \left|#3 \right \rangle}
\newcommand{\braket}[2]{\langle #1|#2\rangle}
\newcommand{\be}{\[}
\newcommand{\br}{\vspace{4mm}}
\newcommand{\bra}[1]{\langle #1|}
\newcommand{\braketbraket}[4]{\langle #1|#2\rangle\langle #3|#4\rangle}
\newcommand{\braop}[2]{\langle #1| \hat #2}
\newcommand{\dd}[1]{ \! \! \!  \mbox{d}#1\ }
\newcommand{\DD}[2]{\frac{\! \! \! \mbox d}{\mbox d #1}#2}
\renewcommand{\det}[1]{\mbox{det}\left(#1\right)}
\newcommand{\ee}{\]} 
\newcommand{\eg}{\textbf{\\  Example: \ \ \ }}
\newcommand{\Imag}[1]{\mbox{Im}\left(#1\right)}
\newcommand{\ket}[1]{|#1\rangle}
\newcommand{\ketbra}[2]{|#1\rangle \langle #2|}
\newcommand{\kp}{\arccos(\frac{\omega - \epsilon}{2t})}
\newcommand{\ldos}{\mbox{L.D.O.S.}}
\renewcommand{\log}[1]{\mbox{log}\left(#1\right)}
\newcommand{\Log}{\mbox{log}}
\newcommand{\Modsq}[1]{\left| #1\right|^2}
\newcommand{\nb}{\textbf{Note: \ \ \ }}
\newcommand{\op}[1]{\hat {#1}}
\newcommand{\opket}[2]{\hat #1 | #2 \rangle}
\newcommand{\occ}{\mbox{Occ. Num.}}
\newcommand{\Real}[1]{\mbox{Re}\left(#1\right)}
\newcommand{\so}{\Rightarrow}
\newcommand{\sol}{\textbf{Solution: \ \ \ }}
\newcommand{\thetafn}[1]{\  \! \theta \left(#1\right)}
\newcommand{\tin}{\int_{-\infty}^{+\infty}\! \! \!\!\!\!\!}
\newcommand{\Tr}[1]{\mbox{Tr}\left(#1\right)}
\newcommand{\kb}{k_B}
\newcommand{\rad}{\mbox{ rad}}
\preprint{APS/123-QED}

\title{Valley Hall Effect and Non-Local Resistance in Locally Gapped Graphene}

\author{Thomas Aktor}
\affiliation{Center for Nanostructured Graphene (CNG), DTU Physics,
	Technical University of Denmark, DK-2800 Kongens Lyngby, Denmark}
\author{Jose H. Garcia}
\affiliation{Catalan Institute of Nanoscience and Nanotechnology (ICN2),
	CSIC and The Barcelona Institute of Science and Technology,
	Campus UAB, Bellaterra, 08193 Barcelona (Cerdanyola del Vall\`es), Spain}
\author{Stephan Roche}
\affiliation{Catalan Institute of Nanoscience and Nanotechnology (ICN2),
	CSIC and The Barcelona Institute of Science and Technology,
	Campus UAB, Bellaterra, 08193 Barcelona (Cerdanyola del Vall\`es), Spain}
\affiliation{ICREA, Instituci\'o Catalana de Recerca i Estudis Avan\c{c}ats,
	08070 Barcelona, Spain}
\author{Antti-Pekka Jauho}
\affiliation{Center for Nanostructured Graphene (CNG), DTU Physics,
	Technical University of Denmark, DK-2800 Kongens Lyngby, Denmark}
\author{Stephen R. Power}
\email{stephen.power@tcd.ie}
\affiliation{School of Physics, Trinity College Dublin, Dublin 2, Ireland}

\date{\today}

\begin{abstract}
We report on the emergence of bulk, valley-polarized currents in graphene-based devices, driven by spatially varying regions of broken sublattice symmetry, and revealed by non-local resistance ($R_\mathrm{NL}$) fingerprints. By using a combination of quantum transport formalisms, giving access to bulk properties as well as multi-terminal device responses, the presence of a \emph{non-uniform local bandgap} is shown to give rise to valley-dependent scattering and a finite Fermi surface contribution to the valley Hall conductivity, related to characteristics of $R_\mathrm{NL}$. These features are robust against disorder and provide a plausible interpretation of controversial experiments in graphene/hBN superlattices. Our findings suggest both an alternative mechanism for the generation of valley Hall effect in graphene, and a route towards valley-dependent electron optics, by materials and device engineering.
\end{abstract}

\maketitle

\label{sec_intro}
Two-dimensional materials, and graphene in particular, are promising valleytronic candidates due to the $K$ and $K^\prime$ valleys at the Dirac points, which can be exploited to encode, transport and process information \cite{schaibley2016valleytronics} and could play a key role in future valley-driven quantum computers \cite{rohling2012universal, PhysRevLett.108.126804, laird2013valley}. 
However, a key hurdle is the absence of external knobs, analogous to magnetic fields and ferromagnetic contacts in spintronics, to generate, manipulate and detect valley-polarized currents \cite{Cresti2016}.
While optoelectronic access to the valley index is possible in certain materials \cite{xiao2012coupledSV, cao2012valley, li2014valleysplitting,PhysRevB.99.115441}, an all-electronic control would be highly desirable for device applications \cite{PhysRevB.96.245410}.
Certain defects or strain deformation fields have been theoretically proposed for achieving valley filtering, but the corresponding experimental implementation remains challenging \cite{rycerz2007valley, garcia-pomar2008valley, fujita2010valley, gunlycke2011valley, chen2014valley, PhysRevB.96.201407, PhysRevApplied.11.044033, levy2010strain, PhysRevB.77.205421, guinea2010energy, vozmediano2010gauge,settnes2016pmandtriaxial, qi2013resonant,settnes2016graphenebub,milovanovic2016strain,settnes2017valleygauge,Zhai2018,Stegmann_2018,wu2018foldwgs, PhysRevB.99.035411, JPSJ.88.083701, PhysRevApplied.11.054019, 2019arXiv190804604T}.
Promising signatures of valley phenomena have instead emerged from non-local resistance ($R_\mathrm{NL}$) measurements in commensurately stacked graphene/hexagonal boron nitride (hBN) systems \cite{gorbachev2014detecting, komatsu2018observation, endo2019topological}.
Large $R_\mathrm{NL}$ signals have been interpreted in terms of an intrinsic valley Hall effect (VHE), driven by bulk Berry curvature \cite{PhysRevLett.114.256601, xiao2007valley, ando2015theory, PhysRevB.94.121408}, and which would be related to a uniform mass term induced by the interaction between graphene and hBN.
Under this mechanism, a quantized valley Hall conductivity, $\sigma_{xy}^v =  \tfrac{2 e^2}{h}$, within the band gap \cite{gorbachev2014detecting,Song2015PNAS,Cresti2016,ando2015theory}, was argued to generate a long-ranged valley current, enhancing $R_\mathrm{NL}$ beyond standard ohmic contributions.
Analogous behaviour has also been studied in bilayer graphene, where the VHE is tunable by an interlayer bias \cite{sui2015gate, shimazaki2015generation}.

However, the interpretation of experimental $R_\mathrm{NL}$ in terms of a bulk-driven VHE has been severely questioned by quantum transport simulations \cite{kirczenow2015valleyNL, Cresti2016, MarmolejoTejada2018}.
Bulk-driven $R_\mathrm{NL}$ signals in fully gapped systems are found to be strongly suppressed beyond evanescent contributions, rendering them fully negligible at experimental scales.
One puzzling issue is that intrinsic valley Hall currents are associated with Fermi sea contributions from Berry curvature hotspots \emph{below} the Fermi energy, whereas within the relevant linear response regime, only Fermi surface contributions should play a role in device measurements \cite{kirczenow2015valleyNL, MarmolejoTejada2018}.
Experimental mapping of current flow further suggested that edge currents may play a role \cite{zhu2017edge}, but recent theoretical \cite{MarmolejoTejada2018, Brown2018, Song2019} and experimental \cite{aharonsteinberg2020longrange} studies cast doubt upon the topological origin of such currents.
Finally, the lattice mismatch between graphene and hBN, which leads to a Moir\'{e} pattern for commensurate structures \cite{gorbachev2014detecting, woods2014commensurate}, also clearly indicates that electrons may not experience a uniform mass term \cite{jung2015origin}, making the interpretation of experimental $R_\mathrm{NL}$ signals in terms of VHE a true and unsolved conundrum.

In this Letter, we demonstrate the emergence of valley-split bulk transport in the absence of a global band gap. 
This phenomenon requires instead the presence of a spatially-varying mass term -- a situation analogous to graphene-hBN heterostructures with commensurate layer alignment.
The valley-polarized current is a consequence of an extrinsic-like valley Hall effect that can be understood by examining the scattering from a single, circular mass dot. 
The exact solution to this case establishes a strong valley dependence in the scattered wavefunction at low energies and a valley-splitting of incoming electronic currents.   
Tight-binding simulations show that this effect is robust for various dot profiles and mass distributions.
The valley Hall conductivity for a periodic array of dots is then calculated using the Kubo-Bastin formalism, and confirms the formation of valley Hall signals and charge neutral currents.
This system manifests a non-zero Fermi surface contribution to the valley Hall conductivity, consistent with the extrinsic contribution and in contrast to the spatially uniform band gap case \cite{Song2015PNAS}.
Using Landauer-Buttiker simulations, we show that scattering-induced valley-polarized currents significantly enhance $R_\mathrm{NL}$ signals, in contrast to the vanishing $R_\mathrm{NL}$  predicted for uniform mass systems. 
Scattering-induced valley splitting, unlike Berry-phase induced deflection, is consistent with standard quantum transport methods, and provides an alternative mechanism to interpret valleytronic phenomena using $R_\mathrm{NL}$ measurements. 
It also suggests a route towards valley engineering by using hBN or other substrates \cite{PhysRevLett.110.046603, woods2014commensurate, jung2015origin, forsythe2018band}, patterned gates \cite{huber2020tunable}, or doping \cite{zhao2011visualizing, lv2012nitrogen, usachov2016large, zabet2014segregation, lawlor2014sublattice, lawlor2014sublattice2, lherbier2013electronic, aktor2016electronic} to generate spatially-varying mass profiles in graphene to direct the flow of valley currents.

\begin{figure}
	\includegraphics[width =0.48\textwidth]{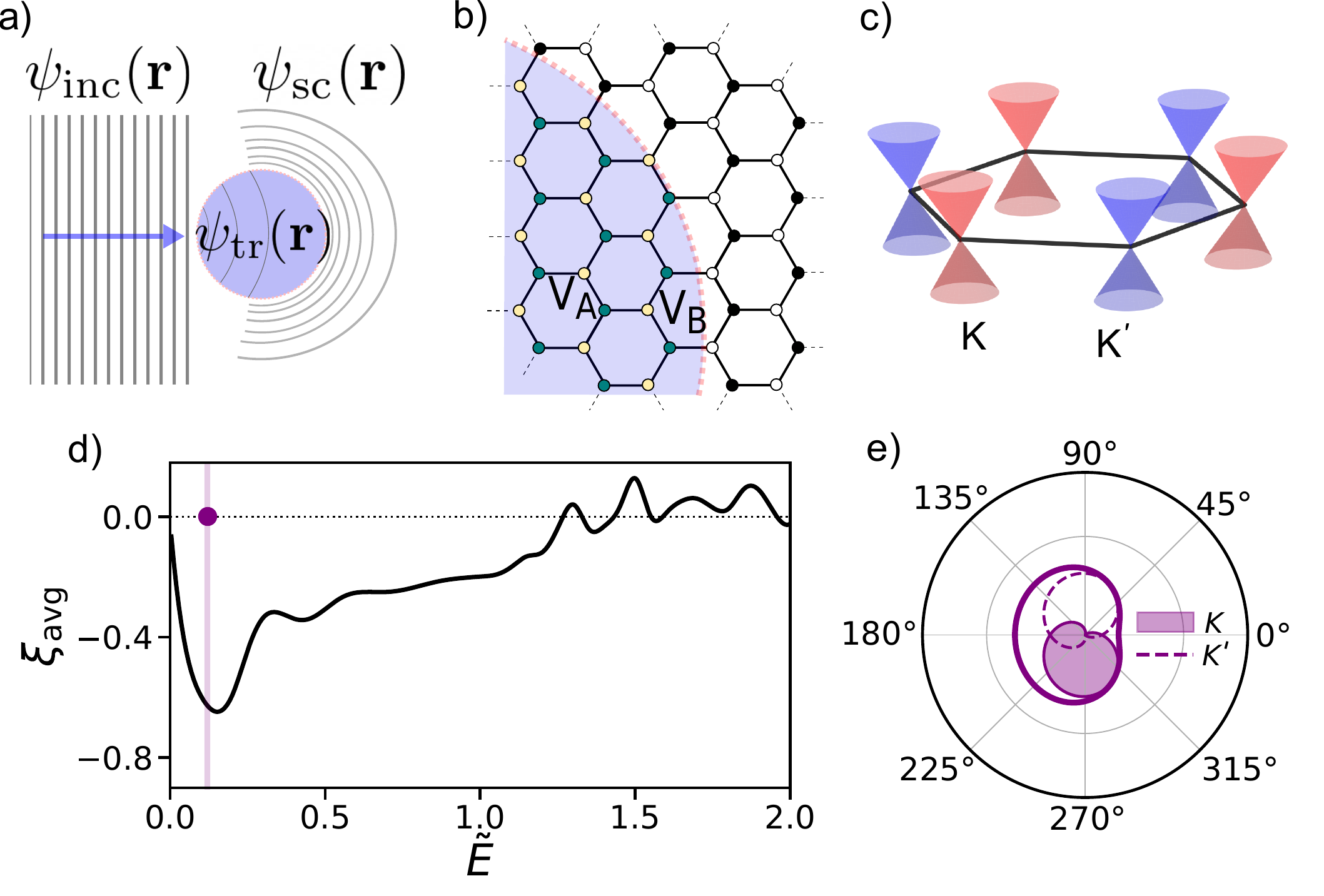}
	\caption{(a) Incoming, scattered and transmitted waves for scattering from a mass dot. (b) $A$ and $B$ sublattice sites in the dot have different onsite potentials. 
	(c) $K$ and $K^\prime$ valleys in the low-energy spectrum of graphene.
	(d) Valley polarization of scattered currents: the peak indicates a strong $K^\prime$ polarization in the $+y$ direction. (e) Total (bold) and individual valley angular scattering profiles in the far-field limit at  $\tilde{E}=0.12$ (purple dot in (d)).}
	\label{fig_Schem1}
\end{figure}


Scattering from a circular mass dot is considered using a Dirac Hamiltonian near the $K$ and $K^\prime$ ($\tau=\pm1$) points 
\begin{equation}
\mathcal{H}_\tau (\mathbf{r}) = 
\hbar v_F\left(
\begin{array}{cc}
\tilde{V}_A (\mathbf{r}) \; & - i\tau \partial_x - \partial_y \\
-i\tau  \partial_x + \partial_y  &  \tilde{V}_B (\mathbf{r}) 
\end{array}
\right) \,,
\label{eqn:DiracH}
\end{equation}
with scaled variables $\tilde{X} \equiv \frac{X}{\hbar v_F}$. 
Here $V_A (\mathbf{r})$, $V_B (\mathbf{r})$ are the A and B sublattice potentials.
Within a dot of radius $R$ we set $V_{A/B} (r<R) = \pm \tfrac{\Delta}{2}$, where the sign depends on the sublattice. 
The mass term $\Delta$ leads to band-gap opening in the range $-\tfrac{\Delta}{2} < E < \tfrac{\Delta}{2}$ in the region where $V_{A/B} \ne 0$.
Following Refs. [\onlinecite{airesf2011unified, heinisch2013, schulz2015electronflow}], we switch to polar coordinates $(r, \phi)$ and consider an incoming ($\psi_{\mathrm{inc}}$), scattered ($\psi_{\mathrm{sc}}$) and  transmitted ($\psi_{\mathrm{tr}}$) waves, as shown in Fig. \ref{fig_Schem1}(a).
$\psi_{\mathrm{inc}}$ and $\psi_{\mathrm{sc}}$ are expanded in terms of angular momentum basis states $m$ using Bessel and Hankel functions respectively, as in the potential dot case \cite{heinisch2013, schulz2015electronflow}.
Inside the dot, $\psi_{\mathrm{tr}}$ is expanded in terms of Bessel functions with sublattice-dependent coefficients \cite{suppmat}.
A closed form expression for the wavefunction is found by enforcing continuity at the dot boundary and solving for the scattering and transmission coefficients, $c^s_m$ and $c^t_m$, for each mode.
The Hamiltonian in Eq. \eqref{eqn:DiracH} is invariant under 
$V_A\leftrightarrow V_B, \psi_1^{\tau} \leftrightarrow \psi_2^{-{\tau}}, \psi_2^{\tau} \leftrightarrow -\psi_1^{-{\tau}}\,,$
so that the $K$ valley result can be used to deduce the $K^\prime$ case.
The local electron density $n=\psi^\dagger\psi$ and particle current $\mathbf{j}=\psi^\dagger\mathbf{\sigma}\psi$ are calculated for 
each valley separately, with the total electronic (valley) quantity given by the sum (difference) of the two valleys.
The local valley polarisation at a point $(r, \phi)$ is the ratio of these quantities  $\xi (r, \phi) \equiv j_{\mathrm{val}}^{\mathrm{sc}} (r, \phi) / j_{\mathrm{tot}}^{\mathrm{sc}} (r, \phi)$, and takes values $-1,0,1$ for fully $K^\prime$ polarized, valley neutral, and fully $K$ polarized currents respectively. 
A figure-of-merit for the valley-splitting efficiency of a dot is 
\begin{equation}
\xi_\mathrm{avg} = \lim_{r\rightarrow\infty} \int_{0}^{\pi} \; \frac{\mathrm{d}\phi} {\pi}  \; \xi^{\mathrm{sc}} (r, \phi) \,,
\end{equation}
\emph{i.e} the far-field limit of the scattered current polarization, averaged over the upper-half plane. 
This quantifies the average valley polarisation of transverse currents, with non-zero $\xi_\mathrm{avg}$ indicating VHE-type behaviour.
We consider a $R=4.5$ dot with $\tilde{\Delta}=2$ so that the band edge inside the dot is at $\tilde{E}=\pm1$, and focus here on valley-dependent effects -- for general scattering properties see \cite{suppmat}.
In Fig \ref{fig_Schem1}(d), $\xi_{\mathrm{avg}}$ takes large negative values as $E$ increases from zero, before decaying to about half its maximum magnitude, and then vanishing at energies above the band edge.
This corresponds to $K^\prime$ electrons preferentially scattering in the $+y$-direction for all band gap energies, with the strongest effect near the gap centre.
The angular dependence of scattering is determined from the radial component of $\mathbf{j}^{\textrm{sc}} (r \to \infty)$.
Fig. \ref{fig_Schem1}(e) plots the total quantity (solid line) and its individual valley contributions (shaded and unshaded areas) at the energy shown by the  dot in Fig. \ref{fig_Schem1}(d)).
$K$ and $K^\prime$ have equal contributions at $\phi=0, \pi$, corresponding to a valley-neutral situation for exact forward- and back-scattering from the dot.
At other angles the two valleys scatter anti-symmetrically with respect to the $x$-axis.
At the energy shown, scattering is largest in the transverse directions so that electrons from different valleys scatter almost entirely to opposite sides of the dot. 

\begin{figure}
	\includegraphics[width =0.49\textwidth]{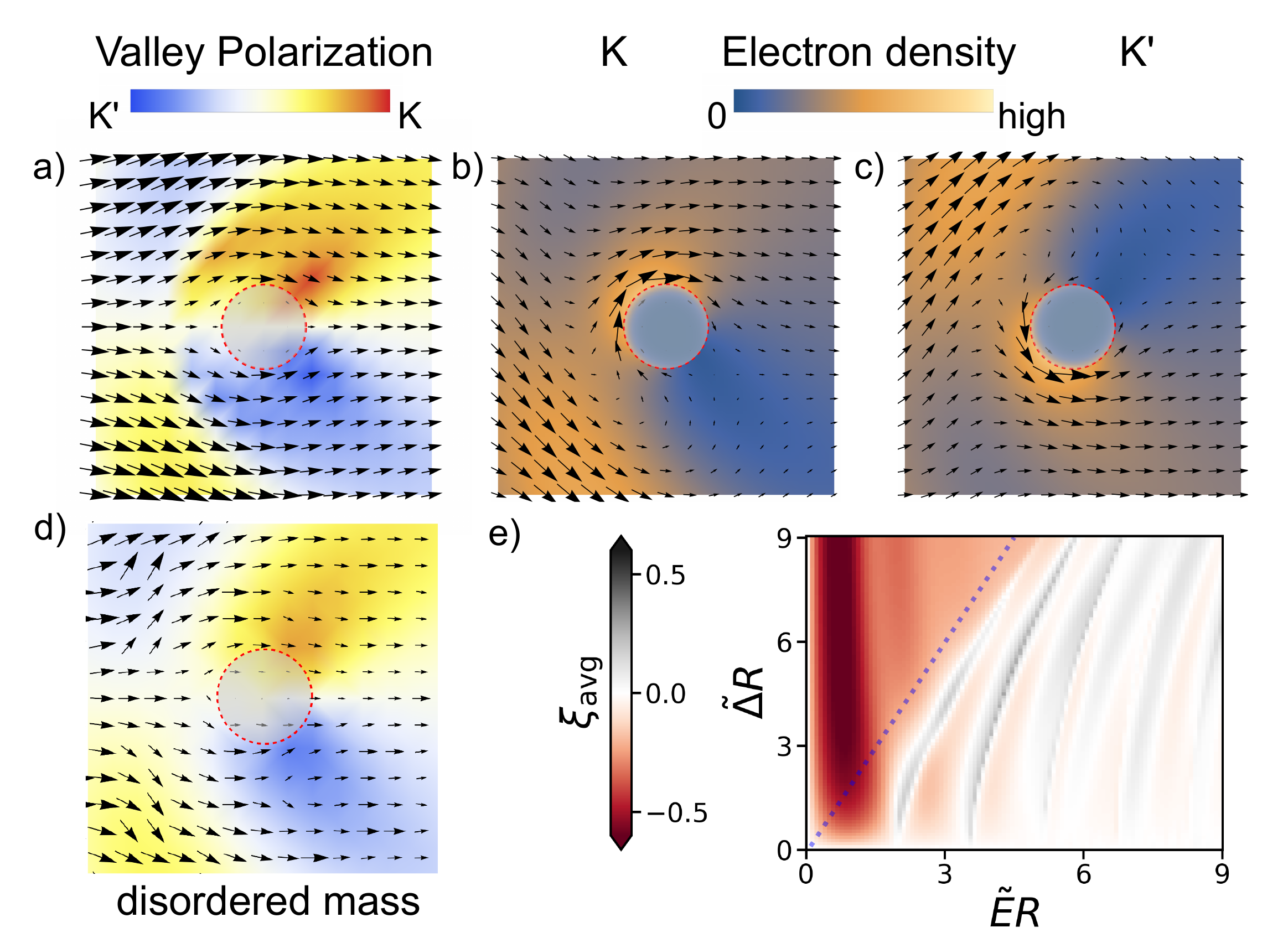}
	\caption{(a) Total  and (b), (c) individual valley current flows near a mass dot at low energy. 
		Electrons from different valleys flow in opposite directions around the dot.
		(d) Similar valley-splitting behaviour for a disordered mass region.
		(e) Phase space map for the average valley polarization $\xi_\mathrm{avg}$ for different $\tilde{E}$ and $\tilde{\Delta}$ values, with the dotted line denoting the band edge.}
	\label{fig_current}
\end{figure}

Fig. \ref{fig_current}(a)-(c) shows the current flow near the dot at this energy, with arrows showing the current magnitude and direction in each case. 
Fig. \ref{fig_current}(a) shows the total charge flow, with the color scale showing valley polarisation, whereas Fig. \ref{fig_current}(b),(c) show the current flow and electron density for each valley separately. 
$K$  and $K^\prime$ electrons flow in opposite directions around the dot, leading to the same antisymmetry around the $x$-axis that was noted for angular scattering.
Valley-polarized currents are observed in the immediate vicinity of the dot in  Fig. \ref{fig_current}(a), but quickly recombine to give a largely valley-neutral forward current.
This is also evident from Fig. \ref{fig_current}(b),(c), where the large $K$ ($K^\prime$ ) current near the top (bottom) of the dot diffuses quickly behind it.
The far-field behaviour emerges from currents further away, namely the $K$ ($K^\prime$ ) current flowing in the bottom (top) left of these panels.

Counterpropagating flows of $K$ and $K^\prime$ currents emerge from the scattering coefficients $c^s_{\bar{m}}(K)$  of the individual modes, whose analytic form is given in \cite{suppmat}.  
The $m$ and $\bar{m} \equiv -m-1$ modes are closely connected, and we find  $c^s_{\bar{m}}(K)=c^s_{m}(K^\prime)$, so that every $K$ valley mode has a corresponding $K^\prime$ mode which contributes with equal magnitude but opposite angular momentum.
At low energies, only $m, \bar{m}=0$ modes contribute, and the clearest valley-splitting is observed.
As $E$ increases, there is a gradual onset of contributions from $m, \bar{m} > 0$ modes, and valley-splitting effects from different modes partially cancel, leading to a decrease in the magnitude of $\xi_{\mathrm{avg}}$ (Fig. \ref{fig_Schem1}(d)). 
However, the presence of higher-order modes maintains the sign of $\xi_{\mathrm{avg}}$ within the gap, so that the deflection direction is consistent. 
This is clear from Fig. \ref{fig_current}e), where we show $\xi_\mathrm{avg}$ across a range of mass strengths and electron energies.
The dotted line denotes the band edge, and a uniform preference for enhanced $K^\prime$ scattering is seen inside the gap, with a weak oscillatory pattern outside due to resonances with bound states.
The strongest valley polarization (dark red region) is at low energies for all sizes and strengths.
For further discussion of the energy dependence, see \cite{suppmat}.

To test the robustness of the valley-splitting predicted above, we perform tight-binding simulations using the patched Green's function approach \cite{settnes2015patched, settnes2016graphenebub}.
A finite dot with $R=10\,\mathrm{nm}$ and $V_{A/B} (r< R) = \pm 0.1|t|$ is embedded in an infinite graphene sheet, and we perform transport calculations using a point probe $250 $ nm away so that  incoming electrons resemble a plane wave, as in Fig. \ref{fig_Schem1}(a).
The system Green's functions \cite{power2011SPAGF}, $G(E)$, and lead broadening term, $\Gamma$, give the spectral density of injected states $A(E)= G  \Gamma G^+ $ and local current flow in real-space.
Projecting $A$ onto the graphene basis $|\psi_{\bm{k}} \rangle$ measures the local distribution of scattered states in k-space, 
$ \rho(\bm{k})=\left <\psi_{\bm{k}} \right|A\left|\psi_{\bm{k}} \right> $, from which the real-space valley polarisation is calculated
$
\xi_{TB}=\left( \sum_{\bm{k} \in K} \rho(\bm{k})-\sum_{\bm{k}\in K^\prime} \rho(\bm{k})\right) \Big/ \sum_{\bm{k}\in K, K^\prime}\rho(\bm{k}) .
$
The analytic polarisation from Fig. \ref{fig_current}(a) is reproduced numerically \cite{suppmat} not only for a perfect dot, but also for non-uniform mass distributions.
Fig. \ref{fig_current}(d) shows a disordered case, where $V_{A/B} \ne 0$ only on 1\% of sites within the dot, but with increased magnitudes to preserve the average mass.
The result, for this and other mass distributions, is almost identical to the analytic, perfect dot prediction.
Insensitivity to the strength, size or composition of mass regions suggests that realisation of valleytronic devices may be easier to achieve using mass dot engineering than strain-based proposals.

\begin{figure}
	\includegraphics[width =0.49\textwidth]{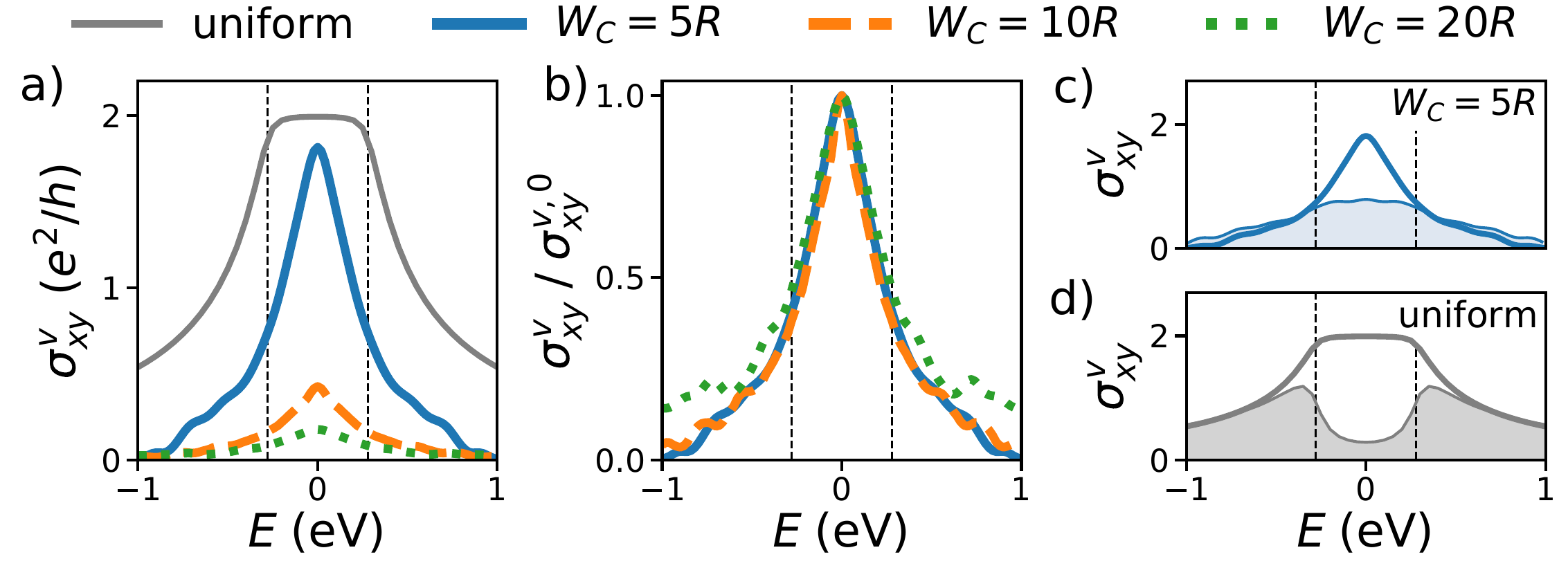}
	\caption{(a) Valley Hall conductivity $\sigma_{xy}^v$, and (b) $\sigma_{xy}^v$ normalized by its peak value, for three different supercell sizes. (c),(d) $\sigma_{xy}^v$ for $W_C=5R$ and uniform mass distributions, together with their Fermi surface contribution (shaded). Dashed vertical lines show the band edges in mass regions.}
	\label{fig_kubo}
\end{figure}

Mass dots induce equal and opposite deflections of $K$ and $K^\prime$ electrons in the $y$ direction, giving rise to a pure valley current in the transverse direction, which is reminiscent of Berry curvature deflections in globally gapped systems. \cite{gorbachev2014detecting,Song2015PNAS,Cresti2016,ando2015theory}.
To determine possible experimental signatures of this phenomenon, we analyze the valley Hall conductivity $\sigma_{xy}^v$ and non-local resistance $R_\mathrm{NL}$ using both the Kubo and Landauer-Buttiker formalisms.
We employ the Kubo-Bastin formula \cite{kubo, kubo-bastin,bruno, Streda1982PRL, garciaPRL,garcia2DMat,garciaNanoLet,settnes2017valleygauge, fan2020} to calculate both the full and Fermi surface contributions to $\sigma_{xy}^v$ \cite{suppmat} for a $R=10$ nm mass dot ($V_{A/B} = \pm 0.1|t|$) placed into one of three different square supercells with side lengths $W_C = 5R, 10R, 20R$. 
This allows us to examine the competition between scattering and Berry curvature effects, which can arise in periodic systems due to a finite concentration $c$ of sites with mass $\Delta$, yielding an effective global mass $ \approx c \Delta$. 
For a global mass, the height of the $\sigma_{xy}^v$ peak remains constant, with wider peaks expected for larger $\Delta_\mathrm{eff}$ (smaller $W_C$). 
However, Fig. \ref{fig_kubo}(a) shows that instead the height decreases with decreasing dot density, while the width is independent of $W_C$, so that the three curves coincide when normalized (Fig. \ref{fig_kubo}(b)).
This suggests a valley-splitting mechanism analogous to an extrinsic spin Hall effect induced by skew-scattering \cite{aires2014skew, aires:extrinsicSHE}.
In this case, the magnitude of $\sigma_{xy}^v$ should vary with the dot density, but with an energy dependence following the scattering profile of a single dot, as observed.
To further substantiate this hypothesis, we examine the Fermi surface (FS) contributions \cite{Streda1982PRL, manchon:kb} to $\sigma_{xy}^v$, which remain finite throughout the band gap for the mass dot system in Fig. \ref{fig_kubo}(c), with $\sigma^{v, \mathrm{FS}}_{xy} \approx 0.4 \,\sigma^{v}_{xy}$ at $E=0$.
For a uniform mass (Fig. \ref{fig_kubo}(d)), this contribution vanishes (aside from broadening effects \cite{suppmat}) in the gap. 
Similar behaviour is found for each $W_C$, demonstrating that a robust Fermi surface contribution emerges in the presence of spatially distributed mass dots.

\begin{figure}
	\includegraphics[width =0.49\textwidth]{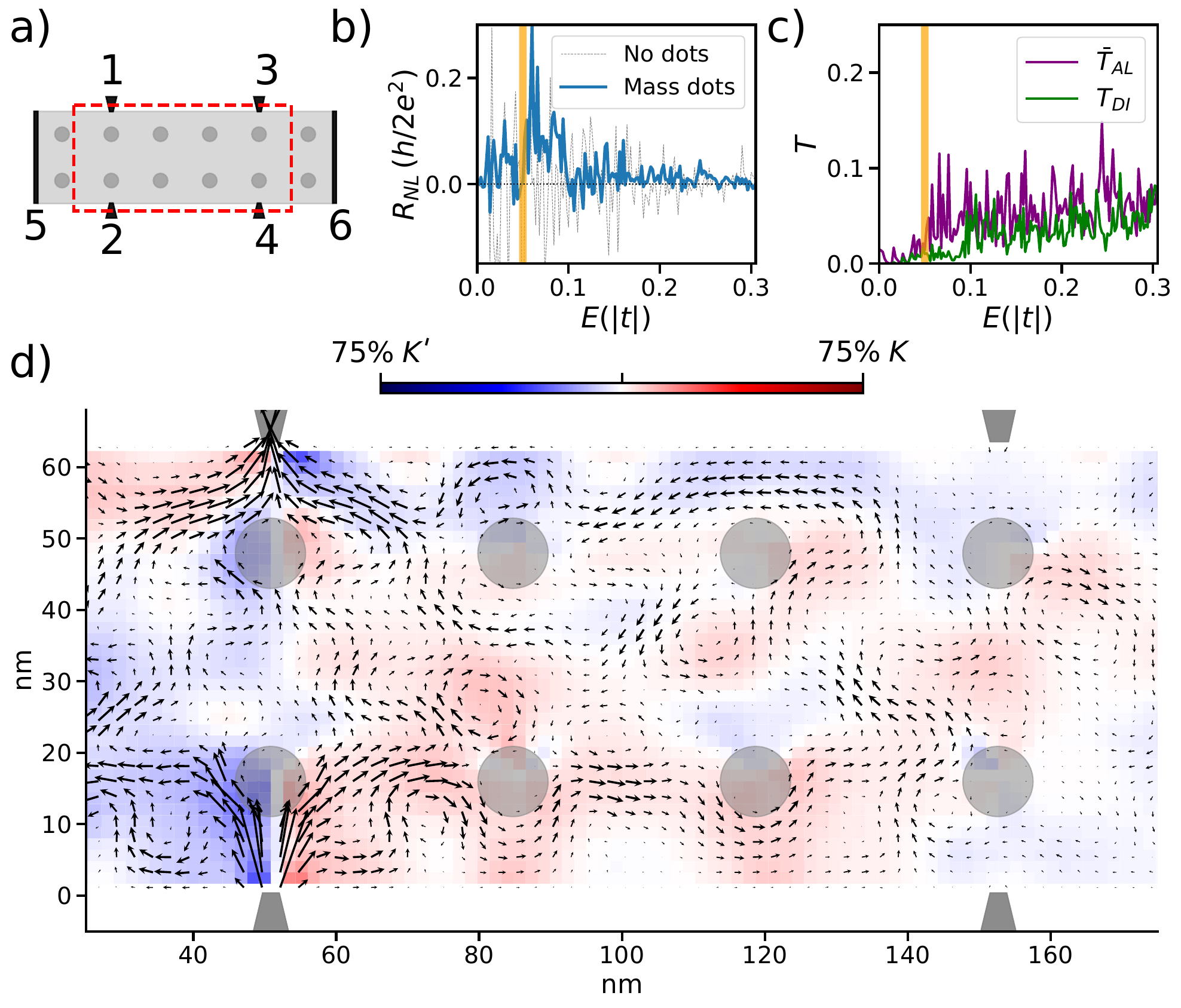}
	\caption{(a) Six-terminal device for $R_\mathrm{NL}$ simulations. 
		(b) $R_\mathrm{NL}$ with and without mass dots, showing a robust positive trend in the presence of dots. 
		(c) Transmission along one edge ($T_{AL}$) and diagonally across the device ($T_{DI}$). 
		(d) Map of local current flow (arrows) and valley polarisation (color) for the energy shown in orange in (b) and (c). $R_\mathrm{NL}$ is mediated by valley-polarized currents in the device.}
	\label{fig_rnl}
\end{figure}

The absence of a global gap enables straightforward $R_\mathrm{NL}$ simulations using Landauer-Buttiker methods.
We consider the six-terminal setup in Fig. \ref{fig_rnl}(a), where $R=5$ nm dots ($V_{A/B} = \pm 0.4|t|$) with periodicity $W_C \sim 6R$ are embedded in a $64$ nm wide zigzag ribbon.
The currents $I_p$ and potentials $V_p$ in each lead $p$ are related by
$
I_p = \frac{2e}{h} \sum \left( T_{qp} V_p - T_{pq}V_q\right),
$
where transmissions $T_{pq}$ are calculated using recursive Green's function techniques \cite{caroli1971direct, Lewenkopf2013, settnes2015patched}.
The potential difference between leads $1$ and $2$ is fixed, and the net current in the remaining leads set to zero.
Solving for the current, $I_1=-I_2$,  and the potentials $V_{3-6}$ yields  
$
R_\mathrm{NL}  = \frac{V_3 - V_4}{I}
$.
In the absence of dots, $R_\mathrm{NL}$ oscillates rapidly around zero, as shown by the dashed line in Fig. \ref{fig_rnl}(b).
These oscillations are due to the finite device size, and emerge from configuration-specific resonances in the pairwise transmissions $T_{pq}$.
Unlike $\sigma^{v}_{xy}$, the $R_\mathrm{NL}$ signal is not mediated solely by Hall effects, but is  sensitive to the device size, edge types, probe placements, and the distances between scatterers and edges. 
Such effects are difficult to remove from the simulations, but average out in experimental-scale systems to give a vanishing contribution to the $R_\mathrm{NL}$ signal.
This represents the current taking a direct path between source and drain, and not probing the non-local region of the device.
With mass dots present (solid blue curve), $R_\mathrm{NL}$ still displays rapid oscillations due to finite size effects, but a wide positive peak feature now emerges at low energies. 
Similar behaviour is seen for different dot sizes and separations, mass strengths, and disorder types, but does not occur in pristine systems or for dots with $V_A = V_B$. 

To understand the origin of the positive $R_\mathrm{NL}$ signal, we analyse the pairwise transmission profiles between probes. 
The sign of $R_\mathrm{NL}$ \cite{suppmat} is a competition between transmissions \emph{along} an edge of the device ($T_{AL}$) and \emph{diagonally} across the device ($T_{DI}$).
Fig \ref{fig_rnl}(c) confirms the positive $R_\mathrm{NL}$ feature corresponds to an enhancement of $T_{AL}$ relative to $T_{DI}$ for our system.
To better interpret the dominant $R_\mathrm{NL}$ signal, we map the local current throughout the device.
In multi-terminal devices, the current flow is not simply an injection at the source and absorption at the drain.
It results from a superposition of currents injected and absorbed by each probe \cite{cresti-currents, nikolic-currents, Lewenkopf2013, Power:GALcommens, vila_2020_nonlocal}.
The arrows in Fig. \ref{fig_rnl}(d) show the current in our system for the energy marked by orange line in Fig. \ref{fig_rnl}(b),(c).
Due to finite size effects, the flow pattern is complex and sensitive to the chosen energy.
Streams and vortices appear, disappear and relocate as the energy is varied.
However, at low energies, strong transverse scattering from the dots leads to indirect current paths which probe large portions of the device.
This is in stark contrast to pristine systems, where more direct paths between source and drain  are observed \cite{suppmat}.
The valley polarization of the current, calculated in the multi-terminal setup, is shown here by color shading.
We note that strong valley splitting at dots near the source and drain contacts, reinforced by subsequent scattering from other dots, leads to valley-polarized currents throughout the entire device.

Electrons from the two valleys tend to flow in opposite directions near the edge, for example, near the bottom edge, $K$ ($K^\prime$) currents flow primarily to the right (left).
This may appear analogous to the quantum spin Hall effect, where electrons of different spin counterpropagate along an edge, but there are fundamental differences.
First the studied system does not have an insulating bulk, but rather strong transverse scattering sources which deflect valley-polarized current towards the device edges.
An important aspect here is that significant current still flows in the regions between dots in the bulk and valley-polarized currents near the edge are not protected or carried by chiral states: they arise due to scattering effects and can be quenched, deflected or change polarization.
However, cumulative scattering from consecutive dots generally acts to reinforce valley polarisation, and to boost $T_{AL}$ by deflecting current along the edge.
This leads to the positive $R_\mathrm{NL}$ feature seen in Fig. \ref{fig_rnl}(b), with similar results found for different device sizes, dot densities, ribbon edges and disorders.
$R_\mathrm{NL}$ signals, mediated by scattering-induced valley splitting, are a general feature of graphene with non-uniform mass distributions.

\paragraph*{Conclusions.}
We calculated the electron scattering from mass dots in graphene, and obtained a clear splitting of electrons according to their valley index. 
This effect is robust over a wide range of dot sizes, mass distributions and electron energies. 
A non-uniform mass distribution, consisting of an array of such dots, gives rise to a valley Hall conductivity which contrasts sharply with that of the uniform mass case and displays a significant Fermi surface contribution.
Furthermore, such arrays give rise to a positive $R_\mathrm{NL}$ feature in Hall bar devices, \emph{without requiring Berry curvature driven, Fermi sea currents in the system bulk.}
Instead, $R_\mathrm{NL}$ signals are driven by cumulative scattering which generates bulk, Fermi-surface, valley-polarized currents and enhances transmission along the device edges. 
While the complicated mass profile in graphene/hBN heterostructures \cite{jung2015origin} makes it difficult to disentangle the different mechanisms at play in this system, non-uniform mass distributions are an ideal platform to further clarify the role of valley currents in the emergence of non-zero $R_\mathrm{NL}$ signals.
Our results suggest mass-based nanostructures are a robust alternative to strained systems to achieve novel valley-dependent electron optics.

\begin{acknowledgments}
S.R.P. acknowledges funding from the Irish Research Council under the Laureate awards programme. 
J.H.G and S.R. acknowledge funding from the European Union Seventh Framework Programme under grant agreement no. 881603 (Graphene Flagship).
ICN2 is funded by the CERCA Programme/Generalitat de Catalunya and supported by the Severo Ochoa programme (MINECO, Grant. No. SEV-2017-0706).
The Center for Nanostructured Graphene (CNG) is sponsored by the Danish National Research Foundation, Project DNRF103.

\end{acknowledgments}


\begin{thebibliography}{85}%
	\makeatletter
	\providecommand \@ifxundefined [1]{%
		\@ifx{#1\undefined}
	}%
	\providecommand \@ifnum [1]{%
		\ifnum #1\expandafter \@firstoftwo
		\else \expandafter \@secondoftwo
		\fi
	}%
	\providecommand \@ifx [1]{%
		\ifx #1\expandafter \@firstoftwo
		\else \expandafter \@secondoftwo
		\fi
	}%
	\providecommand \natexlab [1]{#1}%
	\providecommand \enquote  [1]{``#1''}%
	\providecommand \bibnamefont  [1]{#1}%
	\providecommand \bibfnamefont [1]{#1}%
	\providecommand \citenamefont [1]{#1}%
	\providecommand \href@noop [0]{\@secondoftwo}%
	\providecommand \href [0]{\begingroup \@sanitize@url \@href}%
	\providecommand \@href[1]{\@@startlink{#1}\@@href}%
	\providecommand \@@href[1]{\endgroup#1\@@endlink}%
	\providecommand \@sanitize@url [0]{\catcode `\\12\catcode `\$12\catcode
		`\&12\catcode `\#12\catcode `\^12\catcode `\_12\catcode `\%12\relax}%
	\providecommand \@@startlink[1]{}%
	\providecommand \@@endlink[0]{}%
	\providecommand \url  [0]{\begingroup\@sanitize@url \@url }%
	\providecommand \@url [1]{\endgroup\@href {#1}{\urlprefix }}%
	\providecommand \urlprefix  [0]{URL }%
	\providecommand \Eprint [0]{\href }%
	\providecommand \doibase [0]{http://dx.doi.org/}%
	\providecommand \selectlanguage [0]{\@gobble}%
	\providecommand \bibinfo  [0]{\@secondoftwo}%
	\providecommand \bibfield  [0]{\@secondoftwo}%
	\providecommand \translation [1]{[#1]}%
	\providecommand \BibitemOpen [0]{}%
	\providecommand \bibitemStop [0]{}%
	\providecommand \bibitemNoStop [0]{.\EOS\space}%
	\providecommand \EOS [0]{\spacefactor3000\relax}%
	\providecommand \BibitemShut  [1]{\csname bibitem#1\endcsname}%
	\let\auto@bib@innerbib\@empty
	\bibitem [{\citenamefont {Schaibley}\ \emph {et~al.}(2016)\citenamefont
		{Schaibley}, \citenamefont {Yu}, \citenamefont {Clark}, \citenamefont
		{Rivera}, \citenamefont {Ross}, \citenamefont {Seyler}, \citenamefont {Yao},\
		and\ \citenamefont {Xu}}]{schaibley2016valleytronics}%
	\BibitemOpen
	\bibfield  {author} {\bibinfo {author} {\bibfnamefont {J.~R.}\ \bibnamefont
			{Schaibley}}, \bibinfo {author} {\bibfnamefont {H.}~\bibnamefont {Yu}},
		\bibinfo {author} {\bibfnamefont {G.}~\bibnamefont {Clark}}, \bibinfo
		{author} {\bibfnamefont {P.}~\bibnamefont {Rivera}}, \bibinfo {author}
		{\bibfnamefont {J.~S.}\ \bibnamefont {Ross}}, \bibinfo {author}
		{\bibfnamefont {K.~L.}\ \bibnamefont {Seyler}}, \bibinfo {author}
		{\bibfnamefont {W.}~\bibnamefont {Yao}}, \ and\ \bibinfo {author}
		{\bibfnamefont {X.}~\bibnamefont {Xu}},\ }\href@noop {} {\bibfield  {journal}
		{\bibinfo  {journal} {Nature Reviews Materials}\ }\textbf {\bibinfo {volume}
			{1}},\ \bibinfo {pages} {16055} (\bibinfo {year} {2016})}\BibitemShut
	{NoStop}%
	\bibitem [{\citenamefont {Rohling}\ and\ \citenamefont
		{Burkard}(2012)}]{rohling2012universal}%
	\BibitemOpen
	\bibfield  {author} {\bibinfo {author} {\bibfnamefont {N.}~\bibnamefont
			{Rohling}}\ and\ \bibinfo {author} {\bibfnamefont {G.}~\bibnamefont
			{Burkard}},\ }\href@noop {} {\bibfield  {journal} {\bibinfo  {journal} {New
				Journal of Physics}\ }\textbf {\bibinfo {volume} {14}},\ \bibinfo {pages}
		{083008} (\bibinfo {year} {2012})}\BibitemShut {NoStop}%
	\bibitem [{\citenamefont {Culcer}\ \emph {et~al.}(2012)\citenamefont {Culcer},
		\citenamefont {Saraiva}, \citenamefont {Koiller}, \citenamefont {Hu},\ and\
		\citenamefont {Das~Sarma}}]{PhysRevLett.108.126804}%
	\BibitemOpen
	\bibfield  {author} {\bibinfo {author} {\bibfnamefont {D.}~\bibnamefont
			{Culcer}}, \bibinfo {author} {\bibfnamefont {A.~L.}\ \bibnamefont {Saraiva}},
		\bibinfo {author} {\bibfnamefont {B.}~\bibnamefont {Koiller}}, \bibinfo
		{author} {\bibfnamefont {X.}~\bibnamefont {Hu}}, \ and\ \bibinfo {author}
		{\bibfnamefont {S.}~\bibnamefont {Das~Sarma}},\ }\href@noop {} {\bibfield
		{journal} {\bibinfo  {journal} {Phys. Rev. Lett.}\ }\textbf {\bibinfo
			{volume} {108}},\ \bibinfo {pages} {126804} (\bibinfo {year}
		{2012})}\BibitemShut {NoStop}%
	\bibitem [{\citenamefont {Laird}\ \emph {et~al.}(2013)\citenamefont {Laird},
		\citenamefont {Pei},\ and\ \citenamefont {Kouwenhoven}}]{laird2013valley}%
	\BibitemOpen
	\bibfield  {author} {\bibinfo {author} {\bibfnamefont {E.~A.}\ \bibnamefont
			{Laird}}, \bibinfo {author} {\bibfnamefont {F.}~\bibnamefont {Pei}}, \ and\
		\bibinfo {author} {\bibfnamefont {L.~P.}\ \bibnamefont {Kouwenhoven}},\
	}\href@noop {} {\bibfield  {journal} {\bibinfo  {journal} {Nature
			nanotechnology}\ }\textbf {\bibinfo {volume} {8}},\ \bibinfo {pages} {565}
	(\bibinfo {year} {2013})}\BibitemShut {NoStop}%
\bibitem [{\citenamefont {Cresti}\ \emph {et~al.}(2016)\citenamefont {Cresti},
	\citenamefont {Nikoli{\'{c}}}, \citenamefont {Garc{\'{i}}a},\ and\
	\citenamefont {Roche}}]{Cresti2016}%
\BibitemOpen
\bibfield  {author} {\bibinfo {author} {\bibfnamefont {A.}~\bibnamefont
		{Cresti}}, \bibinfo {author} {\bibfnamefont {B.~K.}\ \bibnamefont
		{Nikoli{\'{c}}}}, \bibinfo {author} {\bibfnamefont {J.~H.}\ \bibnamefont
		{Garc{\'{i}}a}}, \ and\ \bibinfo {author} {\bibfnamefont {S.}~\bibnamefont
		{Roche}},\ }\href@noop {} {\bibfield  {journal} {\bibinfo  {journal} {Riv.
			del Nuovo Cim.}\ }\textbf {\bibinfo {volume} {39}},\ \bibinfo {pages} {587}
	(\bibinfo {year} {2016})}\BibitemShut {NoStop}%
\bibitem [{\citenamefont {Xiao}\ \emph {et~al.}(2012)\citenamefont {Xiao},
	\citenamefont {Liu}, \citenamefont {Feng}, \citenamefont {Xu},\ and\
	\citenamefont {Yao}}]{xiao2012coupledSV}%
\BibitemOpen
\bibfield  {author} {\bibinfo {author} {\bibfnamefont {D.}~\bibnamefont
		{Xiao}}, \bibinfo {author} {\bibfnamefont {G.-B.}\ \bibnamefont {Liu}},
	\bibinfo {author} {\bibfnamefont {W.}~\bibnamefont {Feng}}, \bibinfo {author}
	{\bibfnamefont {X.}~\bibnamefont {Xu}}, \ and\ \bibinfo {author}
	{\bibfnamefont {W.}~\bibnamefont {Yao}},\ }\href@noop {} {\bibfield
	{journal} {\bibinfo  {journal} {Phys. Rev. Lett.}\ }\textbf {\bibinfo
		{volume} {108}},\ \bibinfo {pages} {196802} (\bibinfo {year}
	{2012})}\BibitemShut {NoStop}%
\bibitem [{\citenamefont {Cao}\ \emph {et~al.}(2012)\citenamefont {Cao},
	\citenamefont {Wang}, \citenamefont {Han}, \citenamefont {Ye}, \citenamefont
	{Zhu}, \citenamefont {Shi}, \citenamefont {Niu}, \citenamefont {Tan},
	\citenamefont {Wang}, \citenamefont {Liu} \emph {et~al.}}]{cao2012valley}%
\BibitemOpen
\bibfield  {author} {\bibinfo {author} {\bibfnamefont {T.}~\bibnamefont
		{Cao}}, \bibinfo {author} {\bibfnamefont {G.}~\bibnamefont {Wang}}, \bibinfo
	{author} {\bibfnamefont {W.}~\bibnamefont {Han}}, \bibinfo {author}
	{\bibfnamefont {H.}~\bibnamefont {Ye}}, \bibinfo {author} {\bibfnamefont
		{C.}~\bibnamefont {Zhu}}, \bibinfo {author} {\bibfnamefont {J.}~\bibnamefont
		{Shi}}, \bibinfo {author} {\bibfnamefont {Q.}~\bibnamefont {Niu}}, \bibinfo
	{author} {\bibfnamefont {P.}~\bibnamefont {Tan}}, \bibinfo {author}
	{\bibfnamefont {E.}~\bibnamefont {Wang}}, \bibinfo {author} {\bibfnamefont
		{B.}~\bibnamefont {Liu}},  \emph {et~al.},\ }\href@noop {} {\bibfield
	{journal} {\bibinfo  {journal} {Nature communications}\ }\textbf {\bibinfo
		{volume} {3}},\ \bibinfo {pages} {887} (\bibinfo {year} {2012})}\BibitemShut
{NoStop}%
\bibitem [{\citenamefont {Li}\ \emph {et~al.}(2014)\citenamefont {Li},
	\citenamefont {Ludwig}, \citenamefont {Low}, \citenamefont {Chernikov},
	\citenamefont {Cui}, \citenamefont {Arefe}, \citenamefont {Kim},
	\citenamefont {van~der Zande}, \citenamefont {Rigosi}, \citenamefont {Hill},
	\citenamefont {Kim}, \citenamefont {Hone}, \citenamefont {Li}, \citenamefont
	{Smirnov},\ and\ \citenamefont {Heinz}}]{li2014valleysplitting}%
\BibitemOpen
\bibfield  {author} {\bibinfo {author} {\bibfnamefont {Y.}~\bibnamefont
		{Li}}, \bibinfo {author} {\bibfnamefont {J.}~\bibnamefont {Ludwig}}, \bibinfo
	{author} {\bibfnamefont {T.}~\bibnamefont {Low}}, \bibinfo {author}
	{\bibfnamefont {A.}~\bibnamefont {Chernikov}}, \bibinfo {author}
	{\bibfnamefont {X.}~\bibnamefont {Cui}}, \bibinfo {author} {\bibfnamefont
		{G.}~\bibnamefont {Arefe}}, \bibinfo {author} {\bibfnamefont {Y.~D.}\
		\bibnamefont {Kim}}, \bibinfo {author} {\bibfnamefont {A.~M.}\ \bibnamefont
		{van~der Zande}}, \bibinfo {author} {\bibfnamefont {A.}~\bibnamefont
		{Rigosi}}, \bibinfo {author} {\bibfnamefont {H.~M.}\ \bibnamefont {Hill}},
	\bibinfo {author} {\bibfnamefont {S.~H.}\ \bibnamefont {Kim}}, \bibinfo
	{author} {\bibfnamefont {J.}~\bibnamefont {Hone}}, \bibinfo {author}
	{\bibfnamefont {Z.}~\bibnamefont {Li}}, \bibinfo {author} {\bibfnamefont
		{D.}~\bibnamefont {Smirnov}}, \ and\ \bibinfo {author} {\bibfnamefont
		{T.~F.}\ \bibnamefont {Heinz}},\ }\href@noop {} {\bibfield  {journal}
	{\bibinfo  {journal} {Phys. Rev. Lett.}\ }\textbf {\bibinfo {volume} {113}},\
	\bibinfo {pages} {266804} (\bibinfo {year} {2014})}\BibitemShut {NoStop}%
\bibitem [{\citenamefont {Zhang}\ \emph {et~al.}(2019)\citenamefont {Zhang},
	\citenamefont {Ni}, \citenamefont {Huang}, \citenamefont {Hu},\ and\
	\citenamefont {Liu}}]{PhysRevB.99.115441}%
\BibitemOpen
\bibfield  {author} {\bibinfo {author} {\bibfnamefont {Z.}~\bibnamefont
		{Zhang}}, \bibinfo {author} {\bibfnamefont {X.}~\bibnamefont {Ni}}, \bibinfo
	{author} {\bibfnamefont {H.}~\bibnamefont {Huang}}, \bibinfo {author}
	{\bibfnamefont {L.}~\bibnamefont {Hu}}, \ and\ \bibinfo {author}
	{\bibfnamefont {F.}~\bibnamefont {Liu}},\ }\href@noop {} {\bibfield
	{journal} {\bibinfo  {journal} {Phys. Rev. B}\ }\textbf {\bibinfo {volume}
		{99}},\ \bibinfo {pages} {115441} (\bibinfo {year} {2019})}\BibitemShut
{NoStop}%
\bibitem [{\citenamefont {Ang}\ \emph {et~al.}(2017)\citenamefont {Ang},
	\citenamefont {Yang}, \citenamefont {Zhang}, \citenamefont {Ma},\ and\
	\citenamefont {Ang}}]{PhysRevB.96.245410}%
\BibitemOpen
\bibfield  {author} {\bibinfo {author} {\bibfnamefont {Y.~S.}\ \bibnamefont
		{Ang}}, \bibinfo {author} {\bibfnamefont {S.~A.}\ \bibnamefont {Yang}},
	\bibinfo {author} {\bibfnamefont {C.}~\bibnamefont {Zhang}}, \bibinfo
	{author} {\bibfnamefont {Z.}~\bibnamefont {Ma}}, \ and\ \bibinfo {author}
	{\bibfnamefont {L.~K.}\ \bibnamefont {Ang}},\ }\href@noop {} {\bibfield
	{journal} {\bibinfo  {journal} {Phys. Rev. B}\ }\textbf {\bibinfo {volume}
		{96}},\ \bibinfo {pages} {245410} (\bibinfo {year} {2017})}\BibitemShut
{NoStop}%
\bibitem [{\citenamefont {Rycerz}\ \emph {et~al.}(2007)\citenamefont {Rycerz},
	\citenamefont {Tworzyd{\l}o},\ and\ \citenamefont
	{Beenakker}}]{rycerz2007valley}%
\BibitemOpen
\bibfield  {author} {\bibinfo {author} {\bibfnamefont {A.}~\bibnamefont
		{Rycerz}}, \bibinfo {author} {\bibfnamefont {J.}~\bibnamefont
		{Tworzyd{\l}o}}, \ and\ \bibinfo {author} {\bibfnamefont {C.}~\bibnamefont
		{Beenakker}},\ }\href@noop {} {\bibfield  {journal} {\bibinfo  {journal}
		{Nature Physics}\ }\textbf {\bibinfo {volume} {3}},\ \bibinfo {pages} {172}
	(\bibinfo {year} {2007})}\BibitemShut {NoStop}%
\bibitem [{\citenamefont {Garcia-Pomar}\ \emph {et~al.}(2008)\citenamefont
	{Garcia-Pomar}, \citenamefont {Cortijo},\ and\ \citenamefont
	{Nieto-Vesperinas}}]{garcia-pomar2008valley}%
\BibitemOpen
\bibfield  {author} {\bibinfo {author} {\bibfnamefont {J.~L.}\ \bibnamefont
		{Garcia-Pomar}}, \bibinfo {author} {\bibfnamefont {A.}~\bibnamefont
		{Cortijo}}, \ and\ \bibinfo {author} {\bibfnamefont {M.}~\bibnamefont
		{Nieto-Vesperinas}},\ }\href@noop {} {\bibfield  {journal} {\bibinfo
		{journal} {Phys. Rev. Lett.}\ }\textbf {\bibinfo {volume} {100}},\ \bibinfo
	{pages} {236801} (\bibinfo {year} {2008})}\BibitemShut {NoStop}%
\bibitem [{\citenamefont {Fujita}\ \emph {et~al.}(2010)\citenamefont {Fujita},
	\citenamefont {Jalil},\ and\ \citenamefont {Tan}}]{fujita2010valley}%
\BibitemOpen
\bibfield  {author} {\bibinfo {author} {\bibfnamefont {T.}~\bibnamefont
		{Fujita}}, \bibinfo {author} {\bibfnamefont {M.}~\bibnamefont {Jalil}}, \
	and\ \bibinfo {author} {\bibfnamefont {S.}~\bibnamefont {Tan}},\ }\href@noop
{} {\bibfield  {journal} {\bibinfo  {journal} {Applied Physics Letters}\
	}\textbf {\bibinfo {volume} {97}},\ \bibinfo {pages} {043508} (\bibinfo
	{year} {2010})}\BibitemShut {NoStop}%
\bibitem [{\citenamefont {Gunlycke}\ and\ \citenamefont
	{White}(2011)}]{gunlycke2011valley}%
\BibitemOpen
\bibfield  {author} {\bibinfo {author} {\bibfnamefont {D.}~\bibnamefont
		{Gunlycke}}\ and\ \bibinfo {author} {\bibfnamefont {C.~T.}\ \bibnamefont
		{White}},\ }\href@noop {} {\bibfield  {journal} {\bibinfo  {journal} {Phys.
			Rev. Lett.}\ }\textbf {\bibinfo {volume} {106}},\ \bibinfo {pages} {136806}
	(\bibinfo {year} {2011})}\BibitemShut {NoStop}%
\bibitem [{\citenamefont {Chen}\ \emph {et~al.}(2014)\citenamefont {Chen},
	\citenamefont {Aut\`es}, \citenamefont {Alem}, \citenamefont {Gargiulo},
	\citenamefont {Gautam}, \citenamefont {Linck}, \citenamefont {Kisielowski},
	\citenamefont {Yazyev}, \citenamefont {Louie},\ and\ \citenamefont
	{Zettl}}]{chen2014valley}%
\BibitemOpen
\bibfield  {author} {\bibinfo {author} {\bibfnamefont {J.-H.}\ \bibnamefont
		{Chen}}, \bibinfo {author} {\bibfnamefont {G.}~\bibnamefont {Aut\`es}},
	\bibinfo {author} {\bibfnamefont {N.}~\bibnamefont {Alem}}, \bibinfo {author}
	{\bibfnamefont {F.}~\bibnamefont {Gargiulo}}, \bibinfo {author}
	{\bibfnamefont {A.}~\bibnamefont {Gautam}}, \bibinfo {author} {\bibfnamefont
		{M.}~\bibnamefont {Linck}}, \bibinfo {author} {\bibfnamefont
		{C.}~\bibnamefont {Kisielowski}}, \bibinfo {author} {\bibfnamefont {O.~V.}\
		\bibnamefont {Yazyev}}, \bibinfo {author} {\bibfnamefont {S.~G.}\
		\bibnamefont {Louie}}, \ and\ \bibinfo {author} {\bibfnamefont
		{A.}~\bibnamefont {Zettl}},\ }\href@noop {} {\bibfield  {journal} {\bibinfo
		{journal} {Phys. Rev. B}\ }\textbf {\bibinfo {volume} {89}},\ \bibinfo
	{pages} {121407} (\bibinfo {year} {2014})}\BibitemShut {NoStop}%
\bibitem [{\citenamefont {Asmar}\ and\ \citenamefont
	{Ulloa}(2017)}]{PhysRevB.96.201407}%
\BibitemOpen
\bibfield  {author} {\bibinfo {author} {\bibfnamefont {M.~M.}\ \bibnamefont
		{Asmar}}\ and\ \bibinfo {author} {\bibfnamefont {S.~E.}\ \bibnamefont
		{Ulloa}},\ }\href@noop {} {\bibfield  {journal} {\bibinfo  {journal} {Phys.
			Rev. B}\ }\textbf {\bibinfo {volume} {96}},\ \bibinfo {pages} {201407}
	(\bibinfo {year} {2017})}\BibitemShut {NoStop}%
\bibitem [{\citenamefont {Park}(2019)}]{PhysRevApplied.11.044033}%
\BibitemOpen
\bibfield  {author} {\bibinfo {author} {\bibfnamefont {C.}~\bibnamefont
		{Park}},\ }\href@noop {} {\bibfield  {journal} {\bibinfo  {journal} {Phys.
			Rev. Applied}\ }\textbf {\bibinfo {volume} {11}},\ \bibinfo {pages} {044033}
	(\bibinfo {year} {2019})}\BibitemShut {NoStop}%
\bibitem [{\citenamefont {Levy}\ \emph {et~al.}(2010)\citenamefont {Levy},
	\citenamefont {Burke}, \citenamefont {Meaker}, \citenamefont {Panlasigui},
	\citenamefont {Zettl}, \citenamefont {Guinea}, \citenamefont {Castro~Neto},\
	and\ \citenamefont {Crommie}}]{levy2010strain}%
\BibitemOpen
\bibfield  {author} {\bibinfo {author} {\bibfnamefont {N.}~\bibnamefont
		{Levy}}, \bibinfo {author} {\bibfnamefont {S.}~\bibnamefont {Burke}},
	\bibinfo {author} {\bibfnamefont {K.}~\bibnamefont {Meaker}}, \bibinfo
	{author} {\bibfnamefont {M.}~\bibnamefont {Panlasigui}}, \bibinfo {author}
	{\bibfnamefont {A.}~\bibnamefont {Zettl}}, \bibinfo {author} {\bibfnamefont
		{F.}~\bibnamefont {Guinea}}, \bibinfo {author} {\bibfnamefont {A.~H.}\
		\bibnamefont {Castro~Neto}}, \ and\ \bibinfo {author} {\bibfnamefont
		{M.}~\bibnamefont {Crommie}},\ }\href@noop {} {\bibfield  {journal} {\bibinfo
		{journal} {Science}\ }\textbf {\bibinfo {volume} {329}},\ \bibinfo {pages}
	{544} (\bibinfo {year} {2010})}\BibitemShut {NoStop}%
\bibitem [{\citenamefont {Guinea}\ \emph {et~al.}(2008)\citenamefont {Guinea},
	\citenamefont {Horovitz},\ and\ \citenamefont
	{Le~Doussal}}]{PhysRevB.77.205421}%
\BibitemOpen
\bibfield  {author} {\bibinfo {author} {\bibfnamefont {F.}~\bibnamefont
		{Guinea}}, \bibinfo {author} {\bibfnamefont {B.}~\bibnamefont {Horovitz}}, \
	and\ \bibinfo {author} {\bibfnamefont {P.}~\bibnamefont {Le~Doussal}},\
}\href@noop {} {\bibfield  {journal} {\bibinfo  {journal} {Phys. Rev. B}\
}\textbf {\bibinfo {volume} {77}},\ \bibinfo {pages} {205421} (\bibinfo
{year} {2008})}\BibitemShut {NoStop}%
\bibitem [{\citenamefont {Guinea}\ \emph {et~al.}(2010)\citenamefont {Guinea},
	\citenamefont {Katsnelson},\ and\ \citenamefont {Geim}}]{guinea2010energy}%
\BibitemOpen
\bibfield  {author} {\bibinfo {author} {\bibfnamefont {F.}~\bibnamefont
		{Guinea}}, \bibinfo {author} {\bibfnamefont {M.}~\bibnamefont {Katsnelson}},
	\ and\ \bibinfo {author} {\bibfnamefont {A.}~\bibnamefont {Geim}},\
}\href@noop {} {\bibfield  {journal} {\bibinfo  {journal} {Nature Physics}\
}\textbf {\bibinfo {volume} {6}},\ \bibinfo {pages} {30} (\bibinfo {year}
{2010})}\BibitemShut {NoStop}%
\bibitem [{\citenamefont {Vozmediano}\ \emph {et~al.}(2010)\citenamefont
	{Vozmediano}, \citenamefont {Katsnelson},\ and\ \citenamefont
	{Guinea}}]{vozmediano2010gauge}%
\BibitemOpen
\bibfield  {author} {\bibinfo {author} {\bibfnamefont {M.}~\bibnamefont
		{Vozmediano}}, \bibinfo {author} {\bibfnamefont {M.}~\bibnamefont
		{Katsnelson}}, \ and\ \bibinfo {author} {\bibfnamefont {F.}~\bibnamefont
		{Guinea}},\ }\href@noop {} {\bibfield  {journal} {\bibinfo  {journal}
		{Physics Reports}\ }\textbf {\bibinfo {volume} {496}},\ \bibinfo {pages} {109
	} (\bibinfo {year} {2010})}\BibitemShut {NoStop}%
\bibitem [{\citenamefont {Settnes}\ \emph
	{et~al.}(2016{\natexlab{a}})\citenamefont {Settnes}, \citenamefont {Power},\
	and\ \citenamefont {Jauho}}]{settnes2016pmandtriaxial}%
\BibitemOpen
\bibfield  {author} {\bibinfo {author} {\bibfnamefont {M.}~\bibnamefont
		{Settnes}}, \bibinfo {author} {\bibfnamefont {S.~R.}\ \bibnamefont {Power}},
	\ and\ \bibinfo {author} {\bibfnamefont {A.-P.}\ \bibnamefont {Jauho}},\
}\href@noop {} {\bibfield  {journal} {\bibinfo  {journal} {Phys. Rev. B}\
}\textbf {\bibinfo {volume} {93}},\ \bibinfo {pages} {035456} (\bibinfo
{year} {2016}{\natexlab{a}})}\BibitemShut {NoStop}%
\bibitem [{\citenamefont {Qi}\ \emph {et~al.}(2013)\citenamefont {Qi},
	\citenamefont {Bahamon}, \citenamefont {Pereira}, \citenamefont {Park},
	\citenamefont {Campbell},\ and\ \citenamefont
	{Castro~Neto}}]{qi2013resonant}%
\BibitemOpen
\bibfield  {author} {\bibinfo {author} {\bibfnamefont {Z.}~\bibnamefont
		{Qi}}, \bibinfo {author} {\bibfnamefont {D.}~\bibnamefont {Bahamon}},
	\bibinfo {author} {\bibfnamefont {V.~M.}\ \bibnamefont {Pereira}}, \bibinfo
	{author} {\bibfnamefont {H.~S.}\ \bibnamefont {Park}}, \bibinfo {author}
	{\bibfnamefont {D.}~\bibnamefont {Campbell}}, \ and\ \bibinfo {author}
	{\bibfnamefont {A.~H.}\ \bibnamefont {Castro~Neto}},\ }\href@noop {}
{\bibfield  {journal} {\bibinfo  {journal} {Nano letters}\ }\textbf {\bibinfo
		{volume} {13}},\ \bibinfo {pages} {2692} (\bibinfo {year}
	{2013})}\BibitemShut {NoStop}%
\bibitem [{\citenamefont {Settnes}\ \emph
	{et~al.}(2016{\natexlab{b}})\citenamefont {Settnes}, \citenamefont {Power},
	\citenamefont {Brandbyge},\ and\ \citenamefont
	{Jauho}}]{settnes2016graphenebub}%
\BibitemOpen
\bibfield  {author} {\bibinfo {author} {\bibfnamefont {M.}~\bibnamefont
		{Settnes}}, \bibinfo {author} {\bibfnamefont {S.~R.}\ \bibnamefont {Power}},
	\bibinfo {author} {\bibfnamefont {M.}~\bibnamefont {Brandbyge}}, \ and\
	\bibinfo {author} {\bibfnamefont {A.-P.}\ \bibnamefont {Jauho}},\ }\href@noop
{} {\bibfield  {journal} {\bibinfo  {journal} {Physical review letters}\
	}\textbf {\bibinfo {volume} {117}},\ \bibinfo {pages} {276801} (\bibinfo
	{year} {2016}{\natexlab{b}})}\BibitemShut {NoStop}%
\bibitem [{\citenamefont {Milovanovi{\'c}}\ and\ \citenamefont
	{Peeters}(2016)}]{milovanovic2016strain}%
\BibitemOpen
\bibfield  {author} {\bibinfo {author} {\bibfnamefont {S.}~\bibnamefont
		{Milovanovi{\'c}}}\ and\ \bibinfo {author} {\bibfnamefont {F.}~\bibnamefont
		{Peeters}},\ }\href@noop {} {\bibfield  {journal} {\bibinfo  {journal}
		{Applied Physics Letters}\ }\textbf {\bibinfo {volume} {109}},\ \bibinfo
	{pages} {203108} (\bibinfo {year} {2016})}\BibitemShut {NoStop}%
\bibitem [{\citenamefont {Settnes}\ \emph {et~al.}(2017)\citenamefont
	{Settnes}, \citenamefont {Garcia},\ and\ \citenamefont
	{Roche}}]{settnes2017valleygauge}%
\BibitemOpen
\bibfield  {author} {\bibinfo {author} {\bibfnamefont {M.}~\bibnamefont
		{Settnes}}, \bibinfo {author} {\bibfnamefont {J.~H.}\ \bibnamefont {Garcia}},
	\ and\ \bibinfo {author} {\bibfnamefont {S.}~\bibnamefont {Roche}},\
}\href@noop {} {\bibfield  {journal} {\bibinfo  {journal} {2D Materials}\
}\textbf {\bibinfo {volume} {4}},\ \bibinfo {pages} {031006} (\bibinfo {year}
{2017})}\BibitemShut {NoStop}%
\bibitem [{\citenamefont {Zhai}\ and\ \citenamefont
	{Sandler}(2018)}]{Zhai2018}%
\BibitemOpen
\bibfield  {author} {\bibinfo {author} {\bibfnamefont {D.}~\bibnamefont
		{Zhai}}\ and\ \bibinfo {author} {\bibfnamefont {N.}~\bibnamefont {Sandler}},\
}\href@noop {} {\bibfield  {journal} {\bibinfo  {journal} {Phys. Rev. B}\
}\textbf {\bibinfo {volume} {98}},\ \bibinfo {pages} {165437} (\bibinfo
{year} {2018})}\BibitemShut {NoStop}%
\bibitem [{\citenamefont {Stegmann}\ and\ \citenamefont
	{Szpak}(2018)}]{Stegmann_2018}%
\BibitemOpen
\bibfield  {author} {\bibinfo {author} {\bibfnamefont {T.}~\bibnamefont
		{Stegmann}}\ and\ \bibinfo {author} {\bibfnamefont {N.}~\bibnamefont
		{Szpak}},\ }\href@noop {} {\bibfield  {journal} {\bibinfo  {journal} {2D
			Materials}\ }\textbf {\bibinfo {volume} {6}},\ \bibinfo {pages} {015024}
	(\bibinfo {year} {2018})}\BibitemShut {NoStop}%
\bibitem [{\citenamefont {Wu}\ \emph {et~al.}(2018)\citenamefont {Wu},
	\citenamefont {Zhai}, \citenamefont {Pan}, \citenamefont {Cheng},
	\citenamefont {Taniguchi}, \citenamefont {Watanabe}, \citenamefont
	{Sandler},\ and\ \citenamefont {Bockrath}}]{wu2018foldwgs}%
\BibitemOpen
\bibfield  {author} {\bibinfo {author} {\bibfnamefont {Y.}~\bibnamefont
		{Wu}}, \bibinfo {author} {\bibfnamefont {D.}~\bibnamefont {Zhai}}, \bibinfo
	{author} {\bibfnamefont {C.}~\bibnamefont {Pan}}, \bibinfo {author}
	{\bibfnamefont {B.}~\bibnamefont {Cheng}}, \bibinfo {author} {\bibfnamefont
		{T.}~\bibnamefont {Taniguchi}}, \bibinfo {author} {\bibfnamefont
		{K.}~\bibnamefont {Watanabe}}, \bibinfo {author} {\bibfnamefont
		{N.}~\bibnamefont {Sandler}}, \ and\ \bibinfo {author} {\bibfnamefont
		{M.}~\bibnamefont {Bockrath}},\ }\href@noop {} {\bibfield  {journal}
	{\bibinfo  {journal} {Nano Letters}\ }\textbf {\bibinfo {volume} {18}},\
	\bibinfo {pages} {64} (\bibinfo {year} {2018})}\BibitemShut {NoStop}%
\bibitem [{\citenamefont {Andrade}\ \emph {et~al.}(2019)\citenamefont
	{Andrade}, \citenamefont {Carrillo-Bastos},\ and\ \citenamefont
	{Naumis}}]{PhysRevB.99.035411}%
\BibitemOpen
\bibfield  {author} {\bibinfo {author} {\bibfnamefont {E.}~\bibnamefont
		{Andrade}}, \bibinfo {author} {\bibfnamefont {R.}~\bibnamefont
		{Carrillo-Bastos}}, \ and\ \bibinfo {author} {\bibfnamefont {G.~G.}\
		\bibnamefont {Naumis}},\ }\href@noop {} {\bibfield  {journal} {\bibinfo
		{journal} {Phys. Rev. B}\ }\textbf {\bibinfo {volume} {99}},\ \bibinfo
	{pages} {035411} (\bibinfo {year} {2019})}\BibitemShut {NoStop}%
\bibitem [{\citenamefont {Kariyado}(2019)}]{JPSJ.88.083701}%
\BibitemOpen
\bibfield  {author} {\bibinfo {author} {\bibfnamefont {T.}~\bibnamefont
		{Kariyado}},\ }\href@noop {} {\bibfield  {journal} {\bibinfo  {journal}
		{Journal of the Physical Society of Japan}\ }\textbf {\bibinfo {volume}
		{88}},\ \bibinfo {pages} {083701} (\bibinfo {year} {2019})}\BibitemShut
{NoStop}%
\bibitem [{\citenamefont {McRae}\ \emph {et~al.}(2019)\citenamefont {McRae},
	\citenamefont {Wei},\ and\ \citenamefont
	{Champagne}}]{PhysRevApplied.11.054019}%
\BibitemOpen
\bibfield  {author} {\bibinfo {author} {\bibfnamefont {A.}~\bibnamefont
		{McRae}}, \bibinfo {author} {\bibfnamefont {G.}~\bibnamefont {Wei}}, \ and\
	\bibinfo {author} {\bibfnamefont {A.}~\bibnamefont {Champagne}},\ }\href@noop
{} {\bibfield  {journal} {\bibinfo  {journal} {Phys. Rev. Applied}\ }\textbf
	{\bibinfo {volume} {11}},\ \bibinfo {pages} {054019} (\bibinfo {year}
	{2019})}\BibitemShut {NoStop}%
\bibitem [{\citenamefont {{Torres}}\ \emph {et~al.}(2019)\citenamefont
	{{Torres}}, \citenamefont {{Silva}}, \citenamefont {{de Souza}},
	\citenamefont {{Silva}},\ and\ \citenamefont
	{{Bahamon}}}]{2019arXiv190804604T}%
\BibitemOpen
\bibfield  {author} {\bibinfo {author} {\bibfnamefont {V.}~\bibnamefont
		{{Torres}}}, \bibinfo {author} {\bibfnamefont {P.}~\bibnamefont {{Silva}}},
	\bibinfo {author} {\bibfnamefont {E.~A.~T.}\ \bibnamefont {{de Souza}}},
	\bibinfo {author} {\bibfnamefont {L.~A.}\ \bibnamefont {{Silva}}}, \ and\
	\bibinfo {author} {\bibfnamefont {D.~A.}\ \bibnamefont {{Bahamon}}},\
}\href@noop {} {\bibfield  {journal} {\bibinfo  {journal} {arXiv e-prints}\
	,\ \bibinfo {eid} {arXiv:1908.04604}} (\bibinfo {year} {2019})}\BibitemShut
{NoStop}%
\bibitem [{\citenamefont {Gorbachev}\ \emph {et~al.}(2014)\citenamefont
	{Gorbachev}, \citenamefont {Song}, \citenamefont {Yu}, \citenamefont
	{Kretinin}, \citenamefont {Withers}, \citenamefont {Cao}, \citenamefont
	{Mishchenko}, \citenamefont {Grigorieva}, \citenamefont {Novoselov},
	\citenamefont {Levitov} \emph {et~al.}}]{gorbachev2014detecting}%
\BibitemOpen
\bibfield  {author} {\bibinfo {author} {\bibfnamefont {R.}~\bibnamefont
		{Gorbachev}}, \bibinfo {author} {\bibfnamefont {J.}~\bibnamefont {Song}},
	\bibinfo {author} {\bibfnamefont {G.}~\bibnamefont {Yu}}, \bibinfo {author}
	{\bibfnamefont {A.}~\bibnamefont {Kretinin}}, \bibinfo {author}
	{\bibfnamefont {F.}~\bibnamefont {Withers}}, \bibinfo {author} {\bibfnamefont
		{Y.}~\bibnamefont {Cao}}, \bibinfo {author} {\bibfnamefont {A.}~\bibnamefont
		{Mishchenko}}, \bibinfo {author} {\bibfnamefont {I.}~\bibnamefont
		{Grigorieva}}, \bibinfo {author} {\bibfnamefont {K.}~\bibnamefont
		{Novoselov}}, \bibinfo {author} {\bibfnamefont {L.}~\bibnamefont {Levitov}},
	\emph {et~al.},\ }\href@noop {} {\bibfield  {journal} {\bibinfo  {journal}
		{Science}\ }\textbf {\bibinfo {volume} {346}},\ \bibinfo {pages} {448}
	(\bibinfo {year} {2014})}\BibitemShut {NoStop}%
\bibitem [{\citenamefont {Komatsu}\ \emph {et~al.}(2018)\citenamefont
	{Komatsu}, \citenamefont {Morita}, \citenamefont {Watanabe}, \citenamefont
	{Tsuya}, \citenamefont {Watanabe}, \citenamefont {Taniguchi},\ and\
	\citenamefont {Moriyama}}]{komatsu2018observation}%
\BibitemOpen
\bibfield  {author} {\bibinfo {author} {\bibfnamefont {K.}~\bibnamefont
		{Komatsu}}, \bibinfo {author} {\bibfnamefont {Y.}~\bibnamefont {Morita}},
	\bibinfo {author} {\bibfnamefont {E.}~\bibnamefont {Watanabe}}, \bibinfo
	{author} {\bibfnamefont {D.}~\bibnamefont {Tsuya}}, \bibinfo {author}
	{\bibfnamefont {K.}~\bibnamefont {Watanabe}}, \bibinfo {author}
	{\bibfnamefont {T.}~\bibnamefont {Taniguchi}}, \ and\ \bibinfo {author}
	{\bibfnamefont {S.}~\bibnamefont {Moriyama}},\ }\href@noop {} {\bibfield
	{journal} {\bibinfo  {journal} {Science Advances}\ }\textbf {\bibinfo
		{volume} {4}},\ \bibinfo {pages} {eaaq0194} (\bibinfo {year}
	{2018})}\BibitemShut {NoStop}%
\bibitem [{\citenamefont {Endo}\ \emph {et~al.}(2019)\citenamefont {Endo},
	\citenamefont {Komatsu}, \citenamefont {Iwasaki}, \citenamefont {Watanabe},
	\citenamefont {Tsuya}, \citenamefont {Watanabe}, \citenamefont {Taniguchi},
	\citenamefont {Noguchi}, \citenamefont {Wakayama}, \citenamefont {Morita}
	\emph {et~al.}}]{endo2019topological}%
\BibitemOpen
\bibfield  {author} {\bibinfo {author} {\bibfnamefont {K.}~\bibnamefont
		{Endo}}, \bibinfo {author} {\bibfnamefont {K.}~\bibnamefont {Komatsu}},
	\bibinfo {author} {\bibfnamefont {T.}~\bibnamefont {Iwasaki}}, \bibinfo
	{author} {\bibfnamefont {E.}~\bibnamefont {Watanabe}}, \bibinfo {author}
	{\bibfnamefont {D.}~\bibnamefont {Tsuya}}, \bibinfo {author} {\bibfnamefont
		{K.}~\bibnamefont {Watanabe}}, \bibinfo {author} {\bibfnamefont
		{T.}~\bibnamefont {Taniguchi}}, \bibinfo {author} {\bibfnamefont
		{Y.}~\bibnamefont {Noguchi}}, \bibinfo {author} {\bibfnamefont
		{Y.}~\bibnamefont {Wakayama}}, \bibinfo {author} {\bibfnamefont
		{Y.}~\bibnamefont {Morita}},  \emph {et~al.},\ }\href@noop {} {\bibfield
	{journal} {\bibinfo  {journal} {Applied Physics Letters}\ }\textbf {\bibinfo
		{volume} {114}},\ \bibinfo {pages} {243105} (\bibinfo {year}
	{2019})}\BibitemShut {NoStop}%
\bibitem [{\citenamefont {Lensky}\ \emph {et~al.}(2015)\citenamefont {Lensky},
	\citenamefont {Song}, \citenamefont {Samutpraphoot},\ and\ \citenamefont
	{Levitov}}]{PhysRevLett.114.256601}%
\BibitemOpen
\bibfield  {author} {\bibinfo {author} {\bibfnamefont {Y.~D.}\ \bibnamefont
		{Lensky}}, \bibinfo {author} {\bibfnamefont {J.~C.~W.}\ \bibnamefont {Song}},
	\bibinfo {author} {\bibfnamefont {P.}~\bibnamefont {Samutpraphoot}}, \ and\
	\bibinfo {author} {\bibfnamefont {L.~S.}\ \bibnamefont {Levitov}},\
}\href@noop {} {\bibfield  {journal} {\bibinfo  {journal} {Phys. Rev. Lett.}\
}\textbf {\bibinfo {volume} {114}},\ \bibinfo {pages} {256601} (\bibinfo
{year} {2015})}\BibitemShut {NoStop}%
\bibitem [{\citenamefont {Xiao}\ \emph {et~al.}(2007)\citenamefont {Xiao},
	\citenamefont {Yao},\ and\ \citenamefont {Niu}}]{xiao2007valley}%
\BibitemOpen
\bibfield  {author} {\bibinfo {author} {\bibfnamefont {D.}~\bibnamefont
		{Xiao}}, \bibinfo {author} {\bibfnamefont {W.}~\bibnamefont {Yao}}, \ and\
	\bibinfo {author} {\bibfnamefont {Q.}~\bibnamefont {Niu}},\ }\href@noop {}
{\bibfield  {journal} {\bibinfo  {journal} {Phys. Rev. Lett.}\ }\textbf
	{\bibinfo {volume} {99}},\ \bibinfo {pages} {236809} (\bibinfo {year}
	{2007})}\BibitemShut {NoStop}%
\bibitem [{\citenamefont {Ando}(2015)}]{ando2015theory}%
\BibitemOpen
\bibfield  {author} {\bibinfo {author} {\bibfnamefont {T.}~\bibnamefont
		{Ando}},\ }\href@noop {} {\bibfield  {journal} {\bibinfo  {journal} {Journal
			of the Physical Society of Japan}\ }\textbf {\bibinfo {volume} {84}},\
	\bibinfo {pages} {114705} (\bibinfo {year} {2015})}\BibitemShut {NoStop}%
\bibitem [{\citenamefont {Beconcini}\ \emph {et~al.}(2016)\citenamefont
	{Beconcini}, \citenamefont {Taddei},\ and\ \citenamefont
	{Polini}}]{PhysRevB.94.121408}%
\BibitemOpen
\bibfield  {author} {\bibinfo {author} {\bibfnamefont {M.}~\bibnamefont
		{Beconcini}}, \bibinfo {author} {\bibfnamefont {F.}~\bibnamefont {Taddei}}, \
	and\ \bibinfo {author} {\bibfnamefont {M.}~\bibnamefont {Polini}},\
}\href@noop {} {\bibfield  {journal} {\bibinfo  {journal} {Phys. Rev. B}\
}\textbf {\bibinfo {volume} {94}},\ \bibinfo {pages} {121408} (\bibinfo
{year} {2016})}\BibitemShut {NoStop}%
\bibitem [{\citenamefont {Song}\ \emph {et~al.}(2015)\citenamefont {Song},
	\citenamefont {Samutpraphoot},\ and\ \citenamefont {Levitov}}]{Song2015PNAS}%
\BibitemOpen
\bibfield  {author} {\bibinfo {author} {\bibfnamefont {J.~C.~W.}\
		\bibnamefont {Song}}, \bibinfo {author} {\bibfnamefont {P.}~\bibnamefont
		{Samutpraphoot}}, \ and\ \bibinfo {author} {\bibfnamefont {L.~S.}\
		\bibnamefont {Levitov}},\ }\href@noop {} {\bibfield  {journal} {\bibinfo
		{journal} {Proceedings of the National Academy of Sciences of the United
			States of America}\ }\textbf {\bibinfo {volume} {112}},\ \bibinfo {pages}
	{10879} (\bibinfo {year} {2015})}\BibitemShut {NoStop}%
\bibitem [{\citenamefont {Sui}\ \emph {et~al.}(2015)\citenamefont {Sui},
	\citenamefont {Chen}, \citenamefont {Ma}, \citenamefont {Shan}, \citenamefont
	{Tian}, \citenamefont {Watanabe}, \citenamefont {Taniguchi}, \citenamefont
	{Jin}, \citenamefont {Yao}, \citenamefont {Xiao} \emph
	{et~al.}}]{sui2015gate}%
\BibitemOpen
\bibfield  {author} {\bibinfo {author} {\bibfnamefont {M.}~\bibnamefont
		{Sui}}, \bibinfo {author} {\bibfnamefont {G.}~\bibnamefont {Chen}}, \bibinfo
	{author} {\bibfnamefont {L.}~\bibnamefont {Ma}}, \bibinfo {author}
	{\bibfnamefont {W.-Y.}\ \bibnamefont {Shan}}, \bibinfo {author}
	{\bibfnamefont {D.}~\bibnamefont {Tian}}, \bibinfo {author} {\bibfnamefont
		{K.}~\bibnamefont {Watanabe}}, \bibinfo {author} {\bibfnamefont
		{T.}~\bibnamefont {Taniguchi}}, \bibinfo {author} {\bibfnamefont
		{X.}~\bibnamefont {Jin}}, \bibinfo {author} {\bibfnamefont {W.}~\bibnamefont
		{Yao}}, \bibinfo {author} {\bibfnamefont {D.}~\bibnamefont {Xiao}},  \emph
	{et~al.},\ }\href@noop {} {\bibfield  {journal} {\bibinfo  {journal} {Nature
			Physics}\ }\textbf {\bibinfo {volume} {11}},\ \bibinfo {pages} {1027}
	(\bibinfo {year} {2015})}\BibitemShut {NoStop}%
\bibitem [{\citenamefont {Shimazaki}\ \emph {et~al.}(2015)\citenamefont
	{Shimazaki}, \citenamefont {Yamamoto}, \citenamefont {Borzenets},
	\citenamefont {Watanabe}, \citenamefont {Taniguchi},\ and\ \citenamefont
	{Tarucha}}]{shimazaki2015generation}%
\BibitemOpen
\bibfield  {author} {\bibinfo {author} {\bibfnamefont {Y.}~\bibnamefont
		{Shimazaki}}, \bibinfo {author} {\bibfnamefont {M.}~\bibnamefont {Yamamoto}},
	\bibinfo {author} {\bibfnamefont {I.~V.}\ \bibnamefont {Borzenets}}, \bibinfo
	{author} {\bibfnamefont {K.}~\bibnamefont {Watanabe}}, \bibinfo {author}
	{\bibfnamefont {T.}~\bibnamefont {Taniguchi}}, \ and\ \bibinfo {author}
	{\bibfnamefont {S.}~\bibnamefont {Tarucha}},\ }\href@noop {} {\bibfield
	{journal} {\bibinfo  {journal} {Nature Physics}\ }\textbf {\bibinfo {volume}
		{11}},\ \bibinfo {pages} {1032} (\bibinfo {year} {2015})}\BibitemShut
{NoStop}%
\bibitem [{\citenamefont {Kirczenow}(2015)}]{kirczenow2015valleyNL}%
\BibitemOpen
\bibfield  {author} {\bibinfo {author} {\bibfnamefont {G.}~\bibnamefont
		{Kirczenow}},\ }\href@noop {} {\bibfield  {journal} {\bibinfo  {journal}
		{Phys. Rev. B}\ }\textbf {\bibinfo {volume} {92}},\ \bibinfo {pages} {125425}
	(\bibinfo {year} {2015})}\BibitemShut {NoStop}%
\bibitem [{\citenamefont {Marmolejo-Tejada}\ \emph {et~al.}(2018)\citenamefont
	{Marmolejo-Tejada}, \citenamefont {Garc{\'{\i}}a}, \citenamefont
	{Petrovi{\'{c}}}, \citenamefont {Chang}, \citenamefont {Sheng}, \citenamefont
	{Cresti}, \citenamefont {Plech{\'{a}}{\v{c}}}, \citenamefont {Roche},\ and\
	\citenamefont {Nikoli{\'{c}}}}]{MarmolejoTejada2018}%
\BibitemOpen
\bibfield  {author} {\bibinfo {author} {\bibfnamefont {J.~M.}\ \bibnamefont
		{Marmolejo-Tejada}}, \bibinfo {author} {\bibfnamefont {J.~H.}\ \bibnamefont
		{Garc{\'{\i}}a}}, \bibinfo {author} {\bibfnamefont {M.~D.}\ \bibnamefont
		{Petrovi{\'{c}}}}, \bibinfo {author} {\bibfnamefont {P.-H.}\ \bibnamefont
		{Chang}}, \bibinfo {author} {\bibfnamefont {X.-L.}\ \bibnamefont {Sheng}},
	\bibinfo {author} {\bibfnamefont {A.}~\bibnamefont {Cresti}}, \bibinfo
	{author} {\bibfnamefont {P.}~\bibnamefont {Plech{\'{a}}{\v{c}}}}, \bibinfo
	{author} {\bibfnamefont {S.}~\bibnamefont {Roche}}, \ and\ \bibinfo {author}
	{\bibfnamefont {B.~K.}\ \bibnamefont {Nikoli{\'{c}}}},\ }\href@noop {}
{\bibfield  {journal} {\bibinfo  {journal} {Journal of Physics: Materials}\
	}\textbf {\bibinfo {volume} {1}},\ \bibinfo {pages} {015006} (\bibinfo {year}
	{2018})}\BibitemShut {NoStop}%
\bibitem [{\citenamefont {Zhu}\ \emph {et~al.}(2017)\citenamefont {Zhu},
	\citenamefont {Kretinin}, \citenamefont {Thompson}, \citenamefont {Bandurin},
	\citenamefont {Hu}, \citenamefont {Yu}, \citenamefont {Birkbeck},
	\citenamefont {Mishchenko}, \citenamefont {Vera-Marun}, \citenamefont
	{Watanabe} \emph {et~al.}}]{zhu2017edge}%
\BibitemOpen
\bibfield  {author} {\bibinfo {author} {\bibfnamefont {M.}~\bibnamefont
		{Zhu}}, \bibinfo {author} {\bibfnamefont {A.}~\bibnamefont {Kretinin}},
	\bibinfo {author} {\bibfnamefont {M.~D.}\ \bibnamefont {Thompson}}, \bibinfo
	{author} {\bibfnamefont {D.}~\bibnamefont {Bandurin}}, \bibinfo {author}
	{\bibfnamefont {S.}~\bibnamefont {Hu}}, \bibinfo {author} {\bibfnamefont
		{G.}~\bibnamefont {Yu}}, \bibinfo {author} {\bibfnamefont {J.}~\bibnamefont
		{Birkbeck}}, \bibinfo {author} {\bibfnamefont {A.}~\bibnamefont
		{Mishchenko}}, \bibinfo {author} {\bibfnamefont {I.}~\bibnamefont
		{Vera-Marun}}, \bibinfo {author} {\bibfnamefont {K.}~\bibnamefont
		{Watanabe}},  \emph {et~al.},\ }\href@noop {} {\bibfield  {journal} {\bibinfo
		{journal} {Nature Communications}\ }\textbf {\bibinfo {volume} {8}},\
	\bibinfo {pages} {14552} (\bibinfo {year} {2017})}\BibitemShut {NoStop}%
\bibitem [{\citenamefont {Brown}\ \emph {et~al.}(2018)\citenamefont {Brown},
	\citenamefont {Walet},\ and\ \citenamefont {Guinea}}]{Brown2018}%
\BibitemOpen
\bibfield  {author} {\bibinfo {author} {\bibfnamefont {R.}~\bibnamefont
		{Brown}}, \bibinfo {author} {\bibfnamefont {N.~R.}\ \bibnamefont {Walet}}, \
	and\ \bibinfo {author} {\bibfnamefont {F.}~\bibnamefont {Guinea}},\
}\href@noop {} {\bibfield  {journal} {\bibinfo  {journal} {Phys. Rev. Lett.}\
}\textbf {\bibinfo {volume} {120}},\ \bibinfo {pages} {026802} (\bibinfo
{year} {2018})}\BibitemShut {NoStop}%
\bibitem [{\citenamefont {Song}\ and\ \citenamefont
	{Vignale}(2019)}]{Song2019}%
\BibitemOpen
\bibfield  {author} {\bibinfo {author} {\bibfnamefont {J.~C.~W.}\
		\bibnamefont {Song}}\ and\ \bibinfo {author} {\bibfnamefont {G.}~\bibnamefont
		{Vignale}},\ }\href@noop {} {\bibfield  {journal} {\bibinfo  {journal} {Phys.
			Rev. B}\ }\textbf {\bibinfo {volume} {99}},\ \bibinfo {pages} {235405}
	(\bibinfo {year} {2019})}\BibitemShut {NoStop}%
\bibitem [{\citenamefont {Aharon-Steinberg}\ \emph {et~al.}(2020)\citenamefont
	{Aharon-Steinberg}, \citenamefont {Marguerite}, \citenamefont {Perello},
	\citenamefont {Bagani}, \citenamefont {Holder}, \citenamefont {Myasoedov},
	\citenamefont {Levitov}, \citenamefont {Geim},\ and\ \citenamefont
	{Zeldov}}]{aharonsteinberg2020longrange}%
\BibitemOpen
\bibfield  {author} {\bibinfo {author} {\bibfnamefont {A.}~\bibnamefont
		{Aharon-Steinberg}}, \bibinfo {author} {\bibfnamefont {A.}~\bibnamefont
		{Marguerite}}, \bibinfo {author} {\bibfnamefont {D.~J.}\ \bibnamefont
		{Perello}}, \bibinfo {author} {\bibfnamefont {K.}~\bibnamefont {Bagani}},
	\bibinfo {author} {\bibfnamefont {T.}~\bibnamefont {Holder}}, \bibinfo
	{author} {\bibfnamefont {Y.}~\bibnamefont {Myasoedov}}, \bibinfo {author}
	{\bibfnamefont {L.~S.}\ \bibnamefont {Levitov}}, \bibinfo {author}
	{\bibfnamefont {A.~K.}\ \bibnamefont {Geim}}, \ and\ \bibinfo {author}
	{\bibfnamefont {E.}~\bibnamefont {Zeldov}},\ }\href@noop {} {\enquote
	{\bibinfo {title} {Long-range nontopological edge currents in charge-neutral
			graphene},}\ } (\bibinfo {year} {2020}),\ \Eprint
{http://arxiv.org/abs/2012.02842} {arXiv:2012.02842 [cond-mat.mes-hall]}
\BibitemShut {NoStop}%
\bibitem [{\citenamefont {Woods}\ \emph {et~al.}(2014)\citenamefont {Woods},
	\citenamefont {Britnell}, \citenamefont {Eckmann}, \citenamefont {Ma},
	\citenamefont {Lu}, \citenamefont {Guo}, \citenamefont {Lin}, \citenamefont
	{Yu}, \citenamefont {Cao}, \citenamefont {Gorbachev} \emph
	{et~al.}}]{woods2014commensurate}%
\BibitemOpen
\bibfield  {author} {\bibinfo {author} {\bibfnamefont {C.}~\bibnamefont
		{Woods}}, \bibinfo {author} {\bibfnamefont {L.}~\bibnamefont {Britnell}},
	\bibinfo {author} {\bibfnamefont {A.}~\bibnamefont {Eckmann}}, \bibinfo
	{author} {\bibfnamefont {R.}~\bibnamefont {Ma}}, \bibinfo {author}
	{\bibfnamefont {J.}~\bibnamefont {Lu}}, \bibinfo {author} {\bibfnamefont
		{H.}~\bibnamefont {Guo}}, \bibinfo {author} {\bibfnamefont {X.}~\bibnamefont
		{Lin}}, \bibinfo {author} {\bibfnamefont {G.}~\bibnamefont {Yu}}, \bibinfo
	{author} {\bibfnamefont {Y.}~\bibnamefont {Cao}}, \bibinfo {author}
	{\bibfnamefont {R.}~\bibnamefont {Gorbachev}},  \emph {et~al.},\ }\href@noop
{} {\bibfield  {journal} {\bibinfo  {journal} {Nature physics}\ }\textbf
	{\bibinfo {volume} {10}},\ \bibinfo {pages} {451} (\bibinfo {year}
	{2014})}\BibitemShut {NoStop}%
\bibitem [{\citenamefont {Jung}\ \emph {et~al.}(2015)\citenamefont {Jung},
	\citenamefont {DaSilva}, \citenamefont {MacDonald},\ and\ \citenamefont
	{Adam}}]{jung2015origin}%
\BibitemOpen
\bibfield  {author} {\bibinfo {author} {\bibfnamefont {J.}~\bibnamefont
		{Jung}}, \bibinfo {author} {\bibfnamefont {A.~M.}\ \bibnamefont {DaSilva}},
	\bibinfo {author} {\bibfnamefont {A.~H.}\ \bibnamefont {MacDonald}}, \ and\
	\bibinfo {author} {\bibfnamefont {S.}~\bibnamefont {Adam}},\ }\href@noop {}
{\bibfield  {journal} {\bibinfo  {journal} {Nature Communications}\ }\textbf
	{\bibinfo {volume} {6}},\ \bibinfo {pages} {6308} (\bibinfo {year}
	{2015})}\BibitemShut {NoStop}%
\bibitem [{\citenamefont {Yang}\ \emph {et~al.}(2013)\citenamefont {Yang},
	\citenamefont {Hallal}, \citenamefont {Terrade}, \citenamefont {Waintal},
	\citenamefont {Roche},\ and\ \citenamefont
	{Chshiev}}]{PhysRevLett.110.046603}%
\BibitemOpen
\bibfield  {author} {\bibinfo {author} {\bibfnamefont {H.~X.}\ \bibnamefont
		{Yang}}, \bibinfo {author} {\bibfnamefont {A.}~\bibnamefont {Hallal}},
	\bibinfo {author} {\bibfnamefont {D.}~\bibnamefont {Terrade}}, \bibinfo
	{author} {\bibfnamefont {X.}~\bibnamefont {Waintal}}, \bibinfo {author}
	{\bibfnamefont {S.}~\bibnamefont {Roche}}, \ and\ \bibinfo {author}
	{\bibfnamefont {M.}~\bibnamefont {Chshiev}},\ }\href@noop {} {\bibfield
	{journal} {\bibinfo  {journal} {Phys. Rev. Lett.}\ }\textbf {\bibinfo
		{volume} {110}},\ \bibinfo {pages} {046603} (\bibinfo {year}
	{2013})}\BibitemShut {NoStop}%
\bibitem [{\citenamefont {Forsythe}\ \emph {et~al.}(2018)\citenamefont
	{Forsythe}, \citenamefont {Zhou}, \citenamefont {Watanabe}, \citenamefont
	{Taniguchi}, \citenamefont {Pasupathy}, \citenamefont {Moon}, \citenamefont
	{Koshino}, \citenamefont {Kim},\ and\ \citenamefont
	{Dean}}]{forsythe2018band}%
\BibitemOpen
\bibfield  {author} {\bibinfo {author} {\bibfnamefont {C.}~\bibnamefont
		{Forsythe}}, \bibinfo {author} {\bibfnamefont {X.}~\bibnamefont {Zhou}},
	\bibinfo {author} {\bibfnamefont {K.}~\bibnamefont {Watanabe}}, \bibinfo
	{author} {\bibfnamefont {T.}~\bibnamefont {Taniguchi}}, \bibinfo {author}
	{\bibfnamefont {A.}~\bibnamefont {Pasupathy}}, \bibinfo {author}
	{\bibfnamefont {P.}~\bibnamefont {Moon}}, \bibinfo {author} {\bibfnamefont
		{M.}~\bibnamefont {Koshino}}, \bibinfo {author} {\bibfnamefont
		{P.}~\bibnamefont {Kim}}, \ and\ \bibinfo {author} {\bibfnamefont {C.~R.}\
		\bibnamefont {Dean}},\ }\href@noop {} {\bibfield  {journal} {\bibinfo
		{journal} {Nature nanotechnology}\ }\textbf {\bibinfo {volume} {13}},\
	\bibinfo {pages} {566} (\bibinfo {year} {2018})}\BibitemShut {NoStop}%
\bibitem [{\citenamefont {Huber}\ \emph {et~al.}(2020)\citenamefont {Huber},
	\citenamefont {Liu}, \citenamefont {Chen}, \citenamefont {Drienovsky},
	\citenamefont {Sandner}, \citenamefont {Watanabe}, \citenamefont {Taniguchi},
	\citenamefont {Richter}, \citenamefont {Weiss},\ and\ \citenamefont
	{Eroms}}]{huber2020tunable}%
\BibitemOpen
\bibfield  {author} {\bibinfo {author} {\bibfnamefont {R.}~\bibnamefont
		{Huber}}, \bibinfo {author} {\bibfnamefont {M.-H.}\ \bibnamefont {Liu}},
	\bibinfo {author} {\bibfnamefont {S.-C.}\ \bibnamefont {Chen}}, \bibinfo
	{author} {\bibfnamefont {M.}~\bibnamefont {Drienovsky}}, \bibinfo {author}
	{\bibfnamefont {A.}~\bibnamefont {Sandner}}, \bibinfo {author} {\bibfnamefont
		{K.}~\bibnamefont {Watanabe}}, \bibinfo {author} {\bibfnamefont
		{T.}~\bibnamefont {Taniguchi}}, \bibinfo {author} {\bibfnamefont
		{K.}~\bibnamefont {Richter}}, \bibinfo {author} {\bibfnamefont
		{D.}~\bibnamefont {Weiss}}, \ and\ \bibinfo {author} {\bibfnamefont
		{J.}~\bibnamefont {Eroms}},\ }\href@noop {} {\bibfield  {journal} {\bibinfo
		{journal} {arXiv preprint arXiv:2003.07376}\ } (\bibinfo {year}
	{2020})}\BibitemShut {NoStop}%
\bibitem [{\citenamefont {Zhao}\ \emph {et~al.}(2011)\citenamefont {Zhao},
	\citenamefont {He}, \citenamefont {Rim}, \citenamefont {Schiros},
	\citenamefont {Kim}, \citenamefont {Zhou}, \citenamefont {Guti{\'e}rrez},
	\citenamefont {Chockalingam}, \citenamefont {Arguello}, \citenamefont
	{P{\'a}lov{\'a}} \emph {et~al.}}]{zhao2011visualizing}%
\BibitemOpen
\bibfield  {author} {\bibinfo {author} {\bibfnamefont {L.}~\bibnamefont
		{Zhao}}, \bibinfo {author} {\bibfnamefont {R.}~\bibnamefont {He}}, \bibinfo
	{author} {\bibfnamefont {K.~T.}\ \bibnamefont {Rim}}, \bibinfo {author}
	{\bibfnamefont {T.}~\bibnamefont {Schiros}}, \bibinfo {author} {\bibfnamefont
		{K.~S.}\ \bibnamefont {Kim}}, \bibinfo {author} {\bibfnamefont
		{H.}~\bibnamefont {Zhou}}, \bibinfo {author} {\bibfnamefont {C.}~\bibnamefont
		{Guti{\'e}rrez}}, \bibinfo {author} {\bibfnamefont {S.}~\bibnamefont
		{Chockalingam}}, \bibinfo {author} {\bibfnamefont {C.~J.}\ \bibnamefont
		{Arguello}}, \bibinfo {author} {\bibfnamefont {L.}~\bibnamefont
		{P{\'a}lov{\'a}}},  \emph {et~al.},\ }\href@noop {} {\bibfield  {journal}
	{\bibinfo  {journal} {Science}\ }\textbf {\bibinfo {volume} {333}},\ \bibinfo
	{pages} {999} (\bibinfo {year} {2011})}\BibitemShut {NoStop}%
\bibitem [{\citenamefont {Lv}\ \emph {et~al.}(2012)\citenamefont {Lv},
	\citenamefont {Li}, \citenamefont {Botello-M{\'e}ndez}, \citenamefont
	{Hayashi}, \citenamefont {Wang}, \citenamefont {Berkdemir}, \citenamefont
	{Hao}, \citenamefont {El{\'\i}as}, \citenamefont {Cruz-Silva}, \citenamefont
	{Guti{\'e}rrez} \emph {et~al.}}]{lv2012nitrogen}%
\BibitemOpen
\bibfield  {author} {\bibinfo {author} {\bibfnamefont {R.}~\bibnamefont
		{Lv}}, \bibinfo {author} {\bibfnamefont {Q.}~\bibnamefont {Li}}, \bibinfo
	{author} {\bibfnamefont {A.~R.}\ \bibnamefont {Botello-M{\'e}ndez}}, \bibinfo
	{author} {\bibfnamefont {T.}~\bibnamefont {Hayashi}}, \bibinfo {author}
	{\bibfnamefont {B.}~\bibnamefont {Wang}}, \bibinfo {author} {\bibfnamefont
		{A.}~\bibnamefont {Berkdemir}}, \bibinfo {author} {\bibfnamefont
		{Q.}~\bibnamefont {Hao}}, \bibinfo {author} {\bibfnamefont {A.~L.}\
		\bibnamefont {El{\'\i}as}}, \bibinfo {author} {\bibfnamefont
		{R.}~\bibnamefont {Cruz-Silva}}, \bibinfo {author} {\bibfnamefont {H.~R.}\
		\bibnamefont {Guti{\'e}rrez}},  \emph {et~al.},\ }\href@noop {} {\bibfield
	{journal} {\bibinfo  {journal} {Scientific reports}\ }\textbf {\bibinfo
		{volume} {2}},\ \bibinfo {pages} {586} (\bibinfo {year} {2012})}\BibitemShut
{NoStop}%
\bibitem [{\citenamefont {Usachov}\ \emph {et~al.}(2016)\citenamefont
	{Usachov}, \citenamefont {Fedorov}, \citenamefont {Vilkov}, \citenamefont
	{Petukhov}, \citenamefont {Rybkin}, \citenamefont {Ernst}, \citenamefont
	{Otrokov}, \citenamefont {Chulkov}, \citenamefont {Ogorodnikov},
	\citenamefont {Kuznetsov} \emph {et~al.}}]{usachov2016large}%
\BibitemOpen
\bibfield  {author} {\bibinfo {author} {\bibfnamefont {D.~Y.}\ \bibnamefont
		{Usachov}}, \bibinfo {author} {\bibfnamefont {A.~V.}\ \bibnamefont
		{Fedorov}}, \bibinfo {author} {\bibfnamefont {O.~Y.}\ \bibnamefont {Vilkov}},
	\bibinfo {author} {\bibfnamefont {A.~E.}\ \bibnamefont {Petukhov}}, \bibinfo
	{author} {\bibfnamefont {A.~G.}\ \bibnamefont {Rybkin}}, \bibinfo {author}
	{\bibfnamefont {A.}~\bibnamefont {Ernst}}, \bibinfo {author} {\bibfnamefont
		{M.~M.}\ \bibnamefont {Otrokov}}, \bibinfo {author} {\bibfnamefont {E.~V.}\
		\bibnamefont {Chulkov}}, \bibinfo {author} {\bibfnamefont {I.~I.}\
		\bibnamefont {Ogorodnikov}}, \bibinfo {author} {\bibfnamefont {M.~V.}\
		\bibnamefont {Kuznetsov}},  \emph {et~al.},\ }\href@noop {} {\bibfield
	{journal} {\bibinfo  {journal} {Nano letters}\ }\textbf {\bibinfo {volume}
		{16}},\ \bibinfo {pages} {4535} (\bibinfo {year} {2016})}\BibitemShut
{NoStop}%
\bibitem [{\citenamefont {Zabet-Khosousi}\ \emph {et~al.}(2014)\citenamefont
	{Zabet-Khosousi}, \citenamefont {Zhao}, \citenamefont {P{\'a}lov{\'a}},
	\citenamefont {Hybertsen}, \citenamefont {Reichman}, \citenamefont
	{Pasupathy},\ and\ \citenamefont {Flynn}}]{zabet2014segregation}%
\BibitemOpen
\bibfield  {author} {\bibinfo {author} {\bibfnamefont {A.}~\bibnamefont
		{Zabet-Khosousi}}, \bibinfo {author} {\bibfnamefont {L.}~\bibnamefont
		{Zhao}}, \bibinfo {author} {\bibfnamefont {L.}~\bibnamefont
		{P{\'a}lov{\'a}}}, \bibinfo {author} {\bibfnamefont {M.~S.}\ \bibnamefont
		{Hybertsen}}, \bibinfo {author} {\bibfnamefont {D.~R.}\ \bibnamefont
		{Reichman}}, \bibinfo {author} {\bibfnamefont {A.~N.}\ \bibnamefont
		{Pasupathy}}, \ and\ \bibinfo {author} {\bibfnamefont {G.~W.}\ \bibnamefont
		{Flynn}},\ }\href@noop {} {\bibfield  {journal} {\bibinfo  {journal} {Journal
			of the American Chemical Society}\ }\textbf {\bibinfo {volume} {136}},\
	\bibinfo {pages} {1391} (\bibinfo {year} {2014})}\BibitemShut {NoStop}%
\bibitem [{\citenamefont {Lawlor}\ and\ \citenamefont
	{Ferreira}(2014)}]{lawlor2014sublattice}%
\BibitemOpen
\bibfield  {author} {\bibinfo {author} {\bibfnamefont {J.~A.}\ \bibnamefont
		{Lawlor}}\ and\ \bibinfo {author} {\bibfnamefont {M.~S.}\ \bibnamefont
		{Ferreira}},\ }\href@noop {} {\bibfield  {journal} {\bibinfo  {journal}
		{Beilstein journal of nanotechnology}\ }\textbf {\bibinfo {volume} {5}},\
	\bibinfo {pages} {1210} (\bibinfo {year} {2014})}\BibitemShut {NoStop}%
\bibitem [{\citenamefont {Lawlor}\ \emph {et~al.}(2014)\citenamefont {Lawlor},
	\citenamefont {Gorman}, \citenamefont {Power}, \citenamefont {Bezerra},\ and\
	\citenamefont {Ferreira}}]{lawlor2014sublattice2}%
\BibitemOpen
\bibfield  {author} {\bibinfo {author} {\bibfnamefont {J.~A.}\ \bibnamefont
		{Lawlor}}, \bibinfo {author} {\bibfnamefont {P.~D.}\ \bibnamefont {Gorman}},
	\bibinfo {author} {\bibfnamefont {S.~R.}\ \bibnamefont {Power}}, \bibinfo
	{author} {\bibfnamefont {C.~G.}\ \bibnamefont {Bezerra}}, \ and\ \bibinfo
	{author} {\bibfnamefont {M.~S.}\ \bibnamefont {Ferreira}},\ }\href@noop {}
{\bibfield  {journal} {\bibinfo  {journal} {Carbon}\ }\textbf {\bibinfo
		{volume} {77}},\ \bibinfo {pages} {645} (\bibinfo {year} {2014})}\BibitemShut
{NoStop}%
\bibitem [{\citenamefont {Lherbier}\ \emph {et~al.}(2013)\citenamefont
	{Lherbier}, \citenamefont {Botello-Mendez},\ and\ \citenamefont
	{Charlier}}]{lherbier2013electronic}%
\BibitemOpen
\bibfield  {author} {\bibinfo {author} {\bibfnamefont {A.}~\bibnamefont
		{Lherbier}}, \bibinfo {author} {\bibfnamefont {A.~R.}\ \bibnamefont
		{Botello-Mendez}}, \ and\ \bibinfo {author} {\bibfnamefont {J.-C.}\
		\bibnamefont {Charlier}},\ }\href@noop {} {\bibfield  {journal} {\bibinfo
		{journal} {Nano letters}\ }\textbf {\bibinfo {volume} {13}},\ \bibinfo
	{pages} {1446} (\bibinfo {year} {2013})}\BibitemShut {NoStop}%
\bibitem [{\citenamefont {Aktor}\ \emph {et~al.}(2016)\citenamefont {Aktor},
	\citenamefont {Jauho},\ and\ \citenamefont {Power}}]{aktor2016electronic}%
\BibitemOpen
\bibfield  {author} {\bibinfo {author} {\bibfnamefont {T.}~\bibnamefont
		{Aktor}}, \bibinfo {author} {\bibfnamefont {A.-P.}\ \bibnamefont {Jauho}}, \
	and\ \bibinfo {author} {\bibfnamefont {S.~R.}\ \bibnamefont {Power}},\
}\href@noop {} {\bibfield  {journal} {\bibinfo  {journal} {Physical Review
		B}\ }\textbf {\bibinfo {volume} {93}},\ \bibinfo {pages} {035446} (\bibinfo
{year} {2016})}\BibitemShut {NoStop}%
\bibitem [{\citenamefont {Ferreira}\ \emph {et~al.}(2011)\citenamefont
	{Ferreira}, \citenamefont {Viana-Gomes}, \citenamefont {Nilsson},
	\citenamefont {Mucciolo}, \citenamefont {Peres},\ and\ \citenamefont
	{Castro~Neto}}]{airesf2011unified}%
\BibitemOpen
\bibfield  {author} {\bibinfo {author} {\bibfnamefont {A.}~\bibnamefont
		{Ferreira}}, \bibinfo {author} {\bibfnamefont {J.}~\bibnamefont
		{Viana-Gomes}}, \bibinfo {author} {\bibfnamefont {J.}~\bibnamefont
		{Nilsson}}, \bibinfo {author} {\bibfnamefont {E.~R.}\ \bibnamefont
		{Mucciolo}}, \bibinfo {author} {\bibfnamefont {N.~M.~R.}\ \bibnamefont
		{Peres}}, \ and\ \bibinfo {author} {\bibfnamefont {A.~H.}\ \bibnamefont
		{Castro~Neto}},\ }\href@noop {} {\bibfield  {journal} {\bibinfo  {journal}
		{Phys. Rev. B}\ }\textbf {\bibinfo {volume} {83}},\ \bibinfo {pages} {165402}
	(\bibinfo {year} {2011})}\BibitemShut {NoStop}%
\bibitem [{\citenamefont {Heinisch}\ \emph {et~al.}(2013)\citenamefont
	{Heinisch}, \citenamefont {Bronold},\ and\ \citenamefont
	{Fehske}}]{heinisch2013}%
\BibitemOpen
\bibfield  {author} {\bibinfo {author} {\bibfnamefont {R.~L.}\ \bibnamefont
		{Heinisch}}, \bibinfo {author} {\bibfnamefont {F.~X.}\ \bibnamefont
		{Bronold}}, \ and\ \bibinfo {author} {\bibfnamefont {H.}~\bibnamefont
		{Fehske}},\ }\href@noop {} {\bibfield  {journal} {\bibinfo  {journal}
		{Physical Review B - Condensed Matter and Materials Physics}\ }\textbf
	{\bibinfo {volume} {87}},\ \bibinfo {pages} {1} (\bibinfo {year}
	{2013})}\BibitemShut {NoStop}%
\bibitem [{\citenamefont {Schulz}\ \emph {et~al.}(2015)\citenamefont {Schulz},
	\citenamefont {Heinisch},\ and\ \citenamefont
	{Fehske}}]{schulz2015electronflow}%
\BibitemOpen
\bibfield  {author} {\bibinfo {author} {\bibfnamefont {C.}~\bibnamefont
		{Schulz}}, \bibinfo {author} {\bibfnamefont {R.~L.}\ \bibnamefont
		{Heinisch}}, \ and\ \bibinfo {author} {\bibfnamefont {H.}~\bibnamefont
		{Fehske}},\ }\href@noop {} {\bibfield  {journal} {\bibinfo  {journal}
		{Quantum Matter}\ }\textbf {\bibinfo {volume} {4}},\ \bibinfo {pages} {346}
	(\bibinfo {year} {2015})}\BibitemShut {NoStop}%
\bibitem [{sup()}]{suppmat}%
\BibitemOpen
\href@noop {} {}\bibinfo {note} {See Supplemental Material for the analytic
	solution to the scattering problem, a discussion of mode coefficients and
	details of the Kubo-Bastin calculations}\BibitemShut {NoStop}%
\bibitem [{\citenamefont {Settnes}\ \emph {et~al.}(2015)\citenamefont
	{Settnes}, \citenamefont {Power}, \citenamefont {Lin}, \citenamefont
	{Petersen},\ and\ \citenamefont {Jauho}}]{settnes2015patched}%
\BibitemOpen
\bibfield  {author} {\bibinfo {author} {\bibfnamefont {M.}~\bibnamefont
		{Settnes}}, \bibinfo {author} {\bibfnamefont {S.~R.}\ \bibnamefont {Power}},
	\bibinfo {author} {\bibfnamefont {J.}~\bibnamefont {Lin}}, \bibinfo {author}
	{\bibfnamefont {D.~H.}\ \bibnamefont {Petersen}}, \ and\ \bibinfo {author}
	{\bibfnamefont {A.-P.}\ \bibnamefont {Jauho}},\ }\href@noop {} {\bibfield
	{journal} {\bibinfo  {journal} {Phys. Rev. B}\ }\textbf {\bibinfo {volume}
		{91}},\ \bibinfo {pages} {125408} (\bibinfo {year} {2015})}\BibitemShut
{NoStop}%
\bibitem [{\citenamefont {Power}\ and\ \citenamefont
	{Ferreira}(2011)}]{power2011SPAGF}%
\BibitemOpen
\bibfield  {author} {\bibinfo {author} {\bibfnamefont {S.~R.}\ \bibnamefont
		{Power}}\ and\ \bibinfo {author} {\bibfnamefont {M.~S.}\ \bibnamefont
		{Ferreira}},\ }\href@noop {} {\bibfield  {journal} {\bibinfo  {journal}
		{Phys. Rev. B}\ }\textbf {\bibinfo {volume} {83}},\ \bibinfo {pages} {155432}
	(\bibinfo {year} {2011})}\BibitemShut {NoStop}%
\bibitem [{\citenamefont {Kubo}(1957)}]{kubo}%
\BibitemOpen
\bibfield  {author} {\bibinfo {author} {\bibfnamefont {R.}~\bibnamefont
		{Kubo}},\ }\href@noop {} {\bibfield  {journal} {\bibinfo  {journal} {Journal
			of the Physical Society of Japan}\ }\textbf {\bibinfo {volume} {12}},\
	\bibinfo {pages} {570} (\bibinfo {year} {1957})}\BibitemShut {NoStop}%
\bibitem [{\citenamefont {Bastin}\ \emph {et~al.}(1971)\citenamefont {Bastin},
	\citenamefont {Lewiner}, \citenamefont {Betbeder-Matibet},\ and\
	\citenamefont {Nozieres}}]{kubo-bastin}%
\BibitemOpen
\bibfield  {author} {\bibinfo {author} {\bibfnamefont {A.}~\bibnamefont
		{Bastin}}, \bibinfo {author} {\bibfnamefont {C.}~\bibnamefont {Lewiner}},
	\bibinfo {author} {\bibfnamefont {O.}~\bibnamefont {Betbeder-Matibet}}, \
	and\ \bibinfo {author} {\bibfnamefont {P.}~\bibnamefont {Nozieres}},\
}\href@noop {} {\bibfield  {journal} {\bibinfo  {journal} {Journal of Physics
		and Chemistry of Solids}\ }\textbf {\bibinfo {volume} {32}},\ \bibinfo
{pages} {1811 } (\bibinfo {year} {1971})}\BibitemShut {NoStop}%
\bibitem [{\citenamefont {Cr\'epieux}\ and\ \citenamefont
	{Bruno}(2001)}]{bruno}%
\BibitemOpen
\bibfield  {author} {\bibinfo {author} {\bibfnamefont {A.}~\bibnamefont
		{Cr\'epieux}}\ and\ \bibinfo {author} {\bibfnamefont {P.}~\bibnamefont
		{Bruno}},\ }\href@noop {} {\bibfield  {journal} {\bibinfo  {journal} {Phys.
			Rev. B}\ }\textbf {\bibinfo {volume} {64}},\ \bibinfo {pages} {014416}
	(\bibinfo {year} {2001})}\BibitemShut {NoStop}%
\bibitem [{\citenamefont {Streda}(1982)}]{Streda1982PRL}%
\BibitemOpen
\bibfield  {author} {\bibinfo {author} {\bibfnamefont {P.}~\bibnamefont
		{Streda}},\ }\href@noop {} {\bibfield  {journal} {\bibinfo  {journal}
		{Journal of Physics C: Solid State Physics}\ }\textbf {\bibinfo {volume}
		{15}},\ \bibinfo {pages} {L717} (\bibinfo {year} {1982})}\BibitemShut
{NoStop}%
\bibitem [{\citenamefont {Garc\'{\i}a}\ \emph {et~al.}(2015)\citenamefont
	{Garc\'{\i}a}, \citenamefont {Covaci},\ and\ \citenamefont
	{Rappoport}}]{garciaPRL}%
\BibitemOpen
\bibfield  {author} {\bibinfo {author} {\bibfnamefont {J.~H.}\ \bibnamefont
		{Garc\'{\i}a}}, \bibinfo {author} {\bibfnamefont {L.}~\bibnamefont {Covaci}},
	\ and\ \bibinfo {author} {\bibfnamefont {T.~G.}\ \bibnamefont {Rappoport}},\
}\href@noop {} {\bibfield  {journal} {\bibinfo  {journal} {Phys. Rev. Lett.}\
}\textbf {\bibinfo {volume} {114}},\ \bibinfo {pages} {116602} (\bibinfo
{year} {2015})}\BibitemShut {NoStop}%
\bibitem [{\citenamefont {Garcia}\ and\ \citenamefont
	{Rappoport}(2016)}]{garcia2DMat}%
\BibitemOpen
\bibfield  {author} {\bibinfo {author} {\bibfnamefont {J.~H.}\ \bibnamefont
		{Garcia}}\ and\ \bibinfo {author} {\bibfnamefont {T.~G.}\ \bibnamefont
		{Rappoport}},\ }\href@noop {} {\bibfield  {journal} {\bibinfo  {journal} {2D
			Materials}\ }\textbf {\bibinfo {volume} {3}},\ \bibinfo {pages} {024007}
	(\bibinfo {year} {2016})}\BibitemShut {NoStop}%
\bibitem [{\citenamefont {Garcia}\ \emph {et~al.}(2017)\citenamefont {Garcia},
	\citenamefont {Cummings},\ and\ \citenamefont {Roche}}]{garciaNanoLet}%
\BibitemOpen
\bibfield  {author} {\bibinfo {author} {\bibfnamefont {J.~H.}\ \bibnamefont
		{Garcia}}, \bibinfo {author} {\bibfnamefont {A.~W.}\ \bibnamefont
		{Cummings}}, \ and\ \bibinfo {author} {\bibfnamefont {S.}~\bibnamefont
		{Roche}},\ }\href@noop {} {\bibfield  {journal} {\bibinfo  {journal} {Nano
			letters}\ }\textbf {\bibinfo {volume} {17}},\ \bibinfo {pages} {5078}
	(\bibinfo {year} {2017})}\BibitemShut {NoStop}%
\bibitem [{\citenamefont {Fan}\ \emph {et~al.}(2020)\citenamefont {Fan},
	\citenamefont {Garcia}, \citenamefont {Cummings}, \citenamefont
	{Barrios-Vargas}, \citenamefont {Panhans}, \citenamefont {Harju},
	\citenamefont {Ortmann},\ and\ \citenamefont {Roche}}]{fan2020}%
\BibitemOpen
\bibfield  {author} {\bibinfo {author} {\bibfnamefont {Z.}~\bibnamefont
		{Fan}}, \bibinfo {author} {\bibfnamefont {J.~H.}\ \bibnamefont {Garcia}},
	\bibinfo {author} {\bibfnamefont {A.~W.}\ \bibnamefont {Cummings}}, \bibinfo
	{author} {\bibfnamefont {J.~E.}\ \bibnamefont {Barrios-Vargas}}, \bibinfo
	{author} {\bibfnamefont {M.}~\bibnamefont {Panhans}}, \bibinfo {author}
	{\bibfnamefont {A.}~\bibnamefont {Harju}}, \bibinfo {author} {\bibfnamefont
		{F.}~\bibnamefont {Ortmann}}, \ and\ \bibinfo {author} {\bibfnamefont
		{S.}~\bibnamefont {Roche}},\ }\href@noop {} {\bibfield  {journal} {\bibinfo
		{journal} {Physics Reports}\ } (\bibinfo {year} {2020})}\BibitemShut
{NoStop}%
\bibitem [{\citenamefont {Ferreira}\ \emph {et~al.}(2014)\citenamefont
	{Ferreira}, \citenamefont {Rappoport}, \citenamefont {Cazalilla},\ and\
	\citenamefont {Castro~Neto}}]{aires2014skew}%
\BibitemOpen
\bibfield  {author} {\bibinfo {author} {\bibfnamefont {A.}~\bibnamefont
		{Ferreira}}, \bibinfo {author} {\bibfnamefont {T.~G.}\ \bibnamefont
		{Rappoport}}, \bibinfo {author} {\bibfnamefont {M.~A.}\ \bibnamefont
		{Cazalilla}}, \ and\ \bibinfo {author} {\bibfnamefont {A.~H.}\ \bibnamefont
		{Castro~Neto}},\ }\href@noop {} {\bibfield  {journal} {\bibinfo  {journal}
		{Phys. Rev. Lett.}\ }\textbf {\bibinfo {volume} {112}},\ \bibinfo {pages}
	{066601} (\bibinfo {year} {2014})}\BibitemShut {NoStop}%
\bibitem [{\citenamefont {Milletar\`{\i}}\ and\ \citenamefont
	{Ferreira}(2016)}]{aires:extrinsicSHE}%
\BibitemOpen
\bibfield  {author} {\bibinfo {author} {\bibfnamefont {M.}~\bibnamefont
		{Milletar\`{\i}}}\ and\ \bibinfo {author} {\bibfnamefont {A.}~\bibnamefont
		{Ferreira}},\ }\href@noop {} {\bibfield  {journal} {\bibinfo  {journal}
		{Phys. Rev. B}\ }\textbf {\bibinfo {volume} {94}},\ \bibinfo {pages} {134202}
	(\bibinfo {year} {2016})}\BibitemShut {NoStop}%
\bibitem [{\citenamefont {Bonbien}\ and\ \citenamefont
	{Manchon}(2020)}]{manchon:kb}%
\BibitemOpen
\bibfield  {author} {\bibinfo {author} {\bibfnamefont {V.}~\bibnamefont
		{Bonbien}}\ and\ \bibinfo {author} {\bibfnamefont {A.}~\bibnamefont
		{Manchon}},\ }\href@noop {} {\bibfield  {journal} {\bibinfo  {journal} {Phys.
			Rev. B}\ }\textbf {\bibinfo {volume} {102}},\ \bibinfo {pages} {085113}
	(\bibinfo {year} {2020})}\BibitemShut {NoStop}%
\bibitem [{\citenamefont {Caroli}\ \emph {et~al.}(1971)\citenamefont {Caroli},
	\citenamefont {Combescot}, \citenamefont {Nozieres},\ and\ \citenamefont
	{Saint-James}}]{caroli1971direct}%
\BibitemOpen
\bibfield  {author} {\bibinfo {author} {\bibfnamefont {C.}~\bibnamefont
		{Caroli}}, \bibinfo {author} {\bibfnamefont {R.}~\bibnamefont {Combescot}},
	\bibinfo {author} {\bibfnamefont {P.}~\bibnamefont {Nozieres}}, \ and\
	\bibinfo {author} {\bibfnamefont {D.}~\bibnamefont {Saint-James}},\
}\href@noop {} {\bibfield  {journal} {\bibinfo  {journal} {Journal of Physics
		C: Solid State Physics}\ }\textbf {\bibinfo {volume} {4}},\ \bibinfo {pages}
{916} (\bibinfo {year} {1971})}\BibitemShut {NoStop}%
\bibitem [{\citenamefont {Lewenkopf}\ and\ \citenamefont
	{Mucciolo}(2013)}]{Lewenkopf2013}%
\BibitemOpen
\bibfield  {author} {\bibinfo {author} {\bibfnamefont {C.~H.}\ \bibnamefont
		{Lewenkopf}}\ and\ \bibinfo {author} {\bibfnamefont {E.~R.}\ \bibnamefont
		{Mucciolo}},\ }\href@noop {} {\bibfield  {journal} {\bibinfo  {journal}
		{Journal of Computational Electronics}\ }\textbf {\bibinfo {volume} {12}},\
	\bibinfo {pages} {203} (\bibinfo {year} {2013})}\BibitemShut {NoStop}%
\bibitem [{\citenamefont {Cresti}\ \emph {et~al.}(2003)\citenamefont {Cresti},
	\citenamefont {Farchioni}, \citenamefont {Grosso},\ and\ \citenamefont
	{Parravicini}}]{cresti-currents}%
\BibitemOpen
\bibfield  {author} {\bibinfo {author} {\bibfnamefont {A.}~\bibnamefont
		{Cresti}}, \bibinfo {author} {\bibfnamefont {R.}~\bibnamefont {Farchioni}},
	\bibinfo {author} {\bibfnamefont {G.}~\bibnamefont {Grosso}}, \ and\ \bibinfo
	{author} {\bibfnamefont {G.~P.}\ \bibnamefont {Parravicini}},\ }\href@noop {}
{\bibfield  {journal} {\bibinfo  {journal} {Phys. Rev. B}\ }\textbf {\bibinfo
		{volume} {68}},\ \bibinfo {pages} {075306} (\bibinfo {year}
	{2003})}\BibitemShut {NoStop}%
\bibitem [{\citenamefont {Nikoli\ifmmode~\acute{c}\else \'{c}\fi{}}\ \emph
	{et~al.}(2006)\citenamefont {Nikoli\ifmmode~\acute{c}\else \'{c}\fi{}},
	\citenamefont {Z\^arbo},\ and\ \citenamefont {Souma}}]{nikolic-currents}%
\BibitemOpen
\bibfield  {author} {\bibinfo {author} {\bibfnamefont {B.~K.}\ \bibnamefont
		{Nikoli\ifmmode~\acute{c}\else \'{c}\fi{}}}, \bibinfo {author} {\bibfnamefont
		{L.~P.}\ \bibnamefont {Z\^arbo}}, \ and\ \bibinfo {author} {\bibfnamefont
		{S.}~\bibnamefont {Souma}},\ }\href@noop {} {\bibfield  {journal} {\bibinfo
		{journal} {Phys. Rev. B}\ }\textbf {\bibinfo {volume} {73}},\ \bibinfo
	{pages} {075303} (\bibinfo {year} {2006})}\BibitemShut {NoStop}%
\bibitem [{\citenamefont {Power}\ \emph {et~al.}(2017)\citenamefont {Power},
	\citenamefont {Thomsen}, \citenamefont {Jauho},\ and\ \citenamefont
	{Pedersen}}]{Power:GALcommens}%
\BibitemOpen
\bibfield  {author} {\bibinfo {author} {\bibfnamefont {S.~R.}\ \bibnamefont
		{Power}}, \bibinfo {author} {\bibfnamefont {M.~R.}\ \bibnamefont {Thomsen}},
	\bibinfo {author} {\bibfnamefont {A.-P.}\ \bibnamefont {Jauho}}, \ and\
	\bibinfo {author} {\bibfnamefont {T.~G.}\ \bibnamefont {Pedersen}},\
}\href@noop {} {\bibfield  {journal} {\bibinfo  {journal} {Phys. Rev. B}\
}\textbf {\bibinfo {volume} {96}},\ \bibinfo {pages} {075425} (\bibinfo
{year} {2017})}\BibitemShut {NoStop}%
\bibitem [{\citenamefont {Vila}\ \emph {et~al.}(2020)\citenamefont {Vila},
	\citenamefont {Garcia}, \citenamefont {Cummings}, \citenamefont {Power},
	\citenamefont {Groth}, \citenamefont {Waintal},\ and\ \citenamefont
	{Roche}}]{vila_2020_nonlocal}%
\BibitemOpen
\bibfield  {author} {\bibinfo {author} {\bibfnamefont {M.}~\bibnamefont
		{Vila}}, \bibinfo {author} {\bibfnamefont {J.~H.}\ \bibnamefont {Garcia}},
	\bibinfo {author} {\bibfnamefont {A.~W.}\ \bibnamefont {Cummings}}, \bibinfo
	{author} {\bibfnamefont {S.~R.}\ \bibnamefont {Power}}, \bibinfo {author}
	{\bibfnamefont {C.~W.}\ \bibnamefont {Groth}}, \bibinfo {author}
	{\bibfnamefont {X.}~\bibnamefont {Waintal}}, \ and\ \bibinfo {author}
	{\bibfnamefont {S.}~\bibnamefont {Roche}},\ }\href@noop {} {\bibfield
	{journal} {\bibinfo  {journal} {Phys. Rev. Lett.}\ }\textbf {\bibinfo
		{volume} {124}},\ \bibinfo {pages} {196602} (\bibinfo {year}
	{2020})}\BibitemShut {NoStop}%
\end{thebibliography}
\end{document}